\documentclass{aastex631}
\usepackage{color}
\usepackage{hyperref}
\usepackage{epstopdf}
\usepackage{graphicx}
\usepackage[FIGTOPCAP]{subfigure}
\usepackage{amsmath}

\newcommand{\Lem}{L_{\rm em}}

\newcommand{\gammae}{\gamma_{\rm e}}

\shorttitle{ search GW emission from new-born magnetar }
\shortauthors{Xie et al. }

\begin{document}

\title{Constraining the ellipticity of new-born magnetar with the observational data of Long gamma-ray bursts}

\correspondingauthor{Da-Ming Wei}
\email{dmwei@pmo.ac.cn}

\author{Lang Xie}
\affiliation{Key Laboratory of Dark Matter and Space Astronomy, Purple Mountain Observatory, Chinese Academy of Sciences, Nanjing 210034, China}
\affiliation{School of Astronomy and Space Science, University of Science and Technology of China, Hefei, Anhui 230026, China}

\author{Da-Ming Wei}
\affiliation{Key Laboratory of Dark Matter and Space Astronomy, Purple Mountain Observatory, Chinese Academy of Sciences, Nanjing 210034, China}
\affiliation{School of Astronomy and Space Science, University of Science and Technology of China, Hefei, Anhui 230026, China}

\author{Yun Wang}
\affiliation{Key Laboratory of Dark Matter and Space Astronomy, Purple Mountain Observatory, Chinese Academy of Sciences, Nanjing 210034, China}
\affiliation{School of Astronomy and Space Science, University of Science and Technology of China, Hefei, Anhui 230026, China}

\author{Zhi-Ping Jin}
\affiliation{Key Laboratory of Dark Matter and Space Astronomy, Purple Mountain Observatory, Chinese Academy of Sciences, Nanjing 210034, China}
\affiliation{School of Astronomy and Space Science, University of Science and Technology of China, Hefei, Anhui 230026, China}

\begin{abstract}
  The X-ray plateau emission observed in many Long gamma-ray bursts (LGRBs) has been usually interpreted as the spin-down luminosity of a rapidly spinning, highly magnetized neutron star (millisecond magnetar). If this is true, then the magnetar may emit extended gravitational wave (GW) emission associated with the X-ray plateau due to non-axisymmetric deformation or various stellar oscillations. The advanced LIGO and Virgo detectors have searched for long-duration GW transients for several years, no evidence of GWs from any magnetar has been found until now. In this work, we attempt to search for signature of GW radiation in the electromagnetic observation of 30 LGRBs under the assumption of the magnetar model.
  We utilize the observations of the LGRB plateau to constrain the properties of the new-born magnetar, including the initial spin period $P_0$, diploe magnetic field strength $B_p$ and the ellipticity $\epsilon$. We find that there are some tight relations between magnetar parameters, e.g., $\epsilon \propto B_p^{1.29}$ and $B_p\propto P_0^{1.14}$.
  In addition, we derive the GW strain for magnetar sample via their spin-down processes, and find that the GWs from these objects may not be detectable by the aLIGO and ET detectors. For a rapidly spinning magnetar ($P\sim 1\mbox{ ms}$, $ B \sim10^{15}\mbox{ G}$), the detection horizon for advanced LIGO O5 detector is $\sim 180\mbox{ Mpc}$. The detection of such GW signal associated with the X-ray plateau would be a smoking gun that the central engine of GRB is a magnetar.

\end{abstract}


\section{ Introduction }

The joint detection of the gravitational wave (GW) signal from GW170817 and the corresponding electromagnetic radiation from GRB 170817A \citep{Abbott2017b, Abbott2017a} have opened a new era of multi-messenger astronomy. Although GWs from a binary compact objects merger (including neutron star (NS)-NS, black hole (BH)-BH, NS-BH) have been detected by the advanced LIGO/Virgo detectors \citep{Abbott2016, Abbott2017b, Abbott2020a}, GWs from post-burst magnetar remain unexplored.

It is well known that long GRBs are associated with the core collapse of massive stars. The post-burst remnant could be a
black hole \citep{Popham1999}, or a long-lived  millisecond magnetar \citep{Dai1998a,Dai1998b,Zhang2001}.
Millisecond magnetars are the promising GW emission sources for the advanced LIGO/Virgo detectors,
since it can emit extended GW emission due to non-axisymmetric deformation or various stellar oscillations \citep{Cutler2002,Stella2005,Haskell2008,Dall'Osso2009,Corsi2009,Ciolfi2009,Mastrano2011,de2016,Lasky2016,Suvorov2016,Gao2017,Abbott2019a}.
Exploring such GW emissions from millisecond magnetars can be used to explore the internal physics of NS.

Millisecond magnetars have been considered to be the possible central engine of some GRBs \citep{Usov1992,Thompson1994,Dai1998a,Dai1998b,Zhang2001, Metzger2008,Bucciantini2012,Xie2020}, which can lose rotational energy to launch relativistic jet and then to power the GRB afterglow emission.
Some observed characteristics of GRBs, e.g., long-lived X-ray plateaus and softer extended emission (EE), suggest that at least for some GRBs, their final product may be a millisecond magnetar rather than a black hole no matter whether they are originated from the collapse of massive stars or binary NSs merger \citep{Zhang2001,Metzger2011,Gompertz2013,Rowlinson2013,Gompertz2014,Metzger2014,L2015}.
Magnetar model has been broadly successful in explaining the X-ray plateau and EE component \citep{Zhang2006,Stratta2018,Strang2019,Sarin2020}. The spin-down energy of magnetar could provide the energy source for above two components.
Several other mechanisms that drive the X-ray plateau have also been discussed in previous works.
\citet{Eichler2006} suggested that the plateau may be attributed to a superposition of the decaying tail of the prompt emission and a line of sight that is outside the edge of a jet.
\citet{Beniamini2020} has invoked structured jets to explain plateaus. 
\citet{G2006} proposed that the plateau phase may result from the forward shock radiation during the pre-deceleration, coasting phase in the external medium.

Numerical simulations showed that the magnetar would be born with a strong magnetic field and a rapid rotation which could lead to the stellar deformation or oscillation \citep{Lai1995,Bonazzola1996,Andersson1998, Lindblom1998,Palomba2001,Cutler2002}, and then the magnetar can emit observable GWs \citep{Fan2013, Dall'Osso2009, Doneva2015}.
The newly born magnetar would spin down via a combination of MD torque and GW quadrupole radiations \citep{Shapiro1983, Zhang2001}, and the
GW radiation has a significant effect on the outflows and spin evolution of magnetar, thus the electromagnetic luminosity will exhibit a distinctive evolution feature when GW emission has been taken into account \citep{Dall'Osso2015, Lasky2016}.
The dynamic evolution of magnetar spin-down is related to the braking index.
When the MD radiation dominates the spin-down of the magnetar, the theoretical braking index is $n = 3$. The braking index of $n = 5$ implies that the
GW radiation dominates the magnetar spin-down.
Interestingly, \citet{Fan2013} proposed that identifying this GW radiation signature in observation data of GRBs is possible.
\citet{Lasky2017} has constrained the braking indices of the magnetars by using the observed X-ray light curves.
\citet{L2019} and \citet{Zou2021} systematically calculated the distribution of the braking indices of magnetars in Short and Long GRBs, the results show that the braking indices of a number of GRBs are between $3$ and $5$, implying GW emission already existed in the process of magnetars spin-down.

In this paper, we focus on the spin-down luminosity evolution of millisecond magnetar by considering the effect of GW radiation
and perform a systematic analysis of the LGRBs whose light curves show an X-ray plateau emission followed by a decay segment.
There are total 30 LGRBs in our sample.
Our results show that the spin-down luminosity of 15 LGRBs may include the contribution of GW radiation, and the other 15 LGRBs are dominated by MD radiation.
By modeling the LGRB X-ray plateau light curves, we derive the magnetar parameters and find some tight correlations between these parameters.
Moreover, we derive the GW strain for this magnetar sample via their spin-down processes and estimate the detectability of the resulting GWs.

This paper is set out as follows. We investigated the effect of GW radiation on the spin-down luminosity of magnetar in section 2. In section 3 we describe our sample selection and then fit the LGRB data with the magnetar model. Constraining the properties of the new-born magnetar are presented in section 4. In section 5, we analyze the detectability of the resulting GWs. The discussions and conclusions are presented in section 6.

\section{ Effects of GW radiation on magnetar spin-down}

A newly born magnetar may have a rapid time-varying quadrupole moment due to stellar deformation, so it can emit extended GW emission associated with the GRB X-ray plateau.
Considering the magnetar being spun down via both the MD and GW radiation, the spin evolution could be described as \citep{Shapiro1983, Zhang2001}
\begin{eqnarray}
 \dot{E}_{\rm rot}=-I\Omega \dot{\Omega} &=& L_{\rm EM} + L_{\rm GW} \nonumber \\
 &=& \frac{B^2_{\rm p}R^{6}\Omega^{4}\sin^2\theta}{6c^{3}}+\frac{32GI^{2}\epsilon^{2}\Omega^{6}}{5c^{5}},
\label{eq:spin}
\end{eqnarray}
where $\Omega$ and $\dot{\Omega}$ are the angular frequency and its time derivative, $\theta$ is the tilt angle between the spin axis and magnetic axis, $B_{\rm p}$ is the dipole magnetic field, $\epsilon$ is the ellipticity, $I$ and $R$ are the moment of inertia and the radius of NS, respectively.
In this work, we just adopt the NS with mass $M=1.4\,M_{\sun}$, $I=1\times 10^{45}\mbox{ g cm$^2$}$ and radius $R=10\mbox{ km}$,
and assume that NS has become an orthogonal rotator ($\theta =\pi/2$).

The observed X-ray luminosity $L_{\rm X}= \eta L_{\rm EM}$, where $\eta$ is the efficiency of converting magnetar electromagnetic energy into GRB X-ray emission.
This luminosity is powered by electromagnetic radiation, and the magnetic energy dissipation may occur at a high-efficiency \citep{Zhang2001, Metzger2014}.
Therefore, we adopt the efficiency $\eta\sim0.5$ \citep{Gao2016}.

According to the equation (\ref{eq:spin}), one can derive the spin frequency evolution as
\begin{equation}
\frac{d\Omega}{dt} = -\beta\Omega^3-\gammae\Omega^5, \label{eq:omegadot}
\end{equation}
where $\beta\equiv B^2R^6/6c^3I$, $\gammae\equiv 32GI\varepsilon^2/5c^5$.
The evolution of angular velocity $\Omega$ includes both the MD and GW torques contributions.

From equation (\ref{eq:omegadot}), assuming that the GW emission contribution is negligibly small, i.e. the MD emission dominates the magnetar spin-down, then the electromagnetic luminosity and the spin-down timescale are
\begin{equation}
	L_{\rm EM}=L_{\rm em,0}  \left(1+\frac{t}{\tau_{\rm em}}\right)^{-2},\label{lemd}
\end{equation}
\begin{equation}
	\tau_{\rm em}=\frac{1}{2\beta\Omega_0^2}.\label{tem}
\end{equation}

If GW quadrupole dominates the magnetar spin-down, the electromagnetic luminosity has a distinctive evolution from MD torque-dominated case.
Ignoring the contribution of MD emission in equation (\ref{eq:omegadot}), the electromagnetic luminosity and the spin-down timescale are given by
\begin{equation}
	L_{\rm EM}=L_{\rm em,0}  \left(1+\frac{t}{\tau_{\rm gw}}\right)^{-1}, \label{lgw}
\end{equation}
\begin{equation}
	\tau_{\rm gw}== \frac{1}{2\gammae\Omega_0^4}.\label{tGW}
\end{equation}

\begin{figure}
 \centering
\includegraphics[angle=0,scale=0.7]{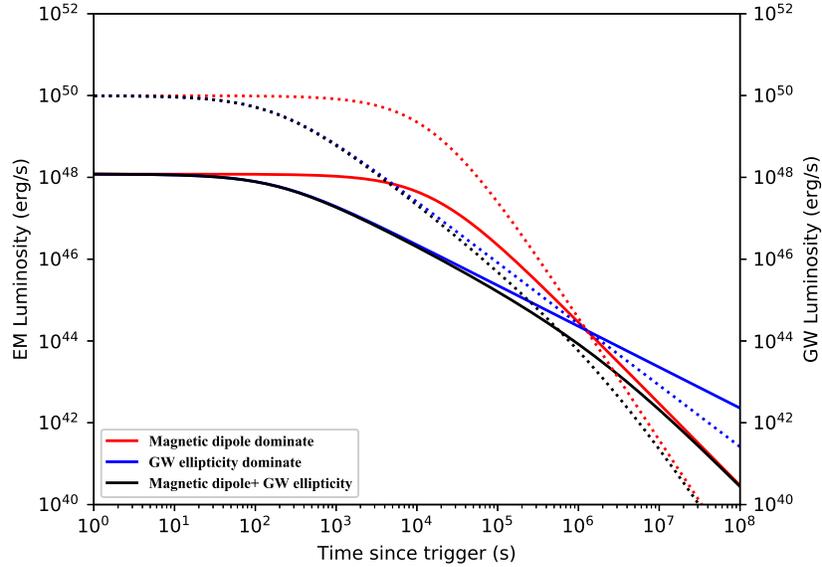}
\caption {The evolution of electromagnetic luminosity (solid line) and gravitational wave luminosity (dashed line) subject to MD radiation and GW radiation losses, with parameters $B=5\times 10^{14}\mbox{G}$, $P=1\mbox{ms}$ and ellipticity $\epsilon=0.001$.
  Red solid and dashed lines show the magnetar spin evolution if only MD radiation losses, and bule solid and dashed lines show the magnetar spin evolution if only GW radiation losses.
  Black solid and dashed lines represent that the spin evolution of magnetar is caused by both MD and GW radiations.}
\label{fig-spindown}
\end{figure}

The solution of the equations (\ref {eq:spin}) and (\ref {eq:omegadot}) with the spin period $P=1\mbox{ ms}$, magnetic field $B=5\times10^{14}\mbox{ G}$ and the ellipticity $\varepsilon = 0.001$, are shown in Fig.\ref{fig-spindown}.
This figure shows the magnetar spin evolution in different scenarios: rotational energy lost by the MD radiation, by the GW radiation, and by the coexistence of MD and GW radiations.
The electromagnetic luminosity generated by the magnetar model evolves as $\Lem\propto(1+t/\tau)^{\alpha}$, where $\tau$ is the spin-down timescale.
The $\alpha$ depends on spin-down torques : $\alpha=-1$ in the case where GW dominates spin-down, $\alpha=-2$ in the case where MD torque dominates and $\alpha$ changes from $-1$ to $-2$ in the case where GW and MD radiations co-dominate spin-down.
From Equation (\ref{eq:omegadot}) we can see that the GW radiation is more efficient than MD radiation at the early time because of the larger rotational angular velocity $\Omega$ at this stage. Therefore GW radiation would dominate magnetar spin-down at the early time and MD radiation would dominate at the late time in a system with GW and MD radiations coexist.
The transition from the GW-dominated phase to the MD-dominated phase is shown as a smooth break, and the decay index of electromagnetic luminosity changes from -1 to -2.
These spin-down luminosity evolution behaviors have been observed in a part of Long and Short GRB light curves, which not only supports the magnetar central engine, but are also used to infer the parameters of magnetars \citep{Rowlinson2013, L2014, Li2018}.

\section{ Search for GW radiation signature in LGRB plateau light curves}

Observations of Long GRBs and their afterglows show that a number of LGRBs are accompanied by an X-ray plateau, which suggests their central engine may be a magnetar.
\citep{Li2018} systematically analyzed the GRB X-ray plateaus, and concluded that the plateau could be powered by the dissipation of magnetar wind.
\citep{L2019} also systematically studied the {\it Swift}/XRT light curves observed during December 2004 - July 2018, and
constrained the braking indices of 45 magnetar candidates.
Our LGRB sample are mainly collected from \citep{Li2018} and \citep{L2019}.

The selection of our magnetar sample should fulfill the following criteria:
Firstly, we select the LGRBs showing a decay segment following the X-ray plateau, and the slope of the decay segment should be between -1 and -2. Those features are consistent with the prediction of the magnetar model, the decay slope -1 and -2 correspond to the situation where GW and MD emission dominated magnetar spin-down, respectively.
Secondly, LGRBs with giant X-ray flares occurring during the spin-down stage are not included in the sample.
These flares are considered to be the re-activities of the GRB central engine.
By using these criteria for sample selection, 30 LGRBs meet our requirements.
For LGRBs without redshift $z$ measurement, $z=1$ is adopted.
We make the K-correction for X-ray data of the LGRB sample \citep{Bloom2001}.
The X-ray data are obtained from the {\it Swift} data archive \citep{Evans2007,Evans2009}. \footnote{http://www.swift.ac.uk/burst\_analyser/}.

The main purpose of this paper is to search for the signature of GW radiation in the electromagnetic observations of LGRBs and place a constraint on the parameters of magnetar. We consider three scenarios, e.g., MD dominates spin-down (MD model), GW dominates spin-down (GW model), MD and GW co-dominates spin-down (hybrid model), to fit the X-ray light curves of our magnetar sample.

We employ Markov Chain Monte Carlo (MCMC) method and emcee Python package to derive the best-fitting model and posterior parameters \citep{Foreman-Mackey2013}. The free parameters in the model include: initial spin period $P_0$, diploe magnetic field strength $B_p$ and the ellipticity $\epsilon$. The prior is set to a log uniform and sufficiently large interval, e.g., $P_0\in [ 0.5-20\mbox{ms}]$, $B_p\in[10^{13}-10^{16}\mbox{ G}]$, $\epsilon\in[10^{-5}-10^{-2}]$.
Based on the value of $\chi^{2}$/dof, we determine the best-fitting model for each burst.

\begin{figure*}
\centering

\includegraphics[width=0.35\textwidth]{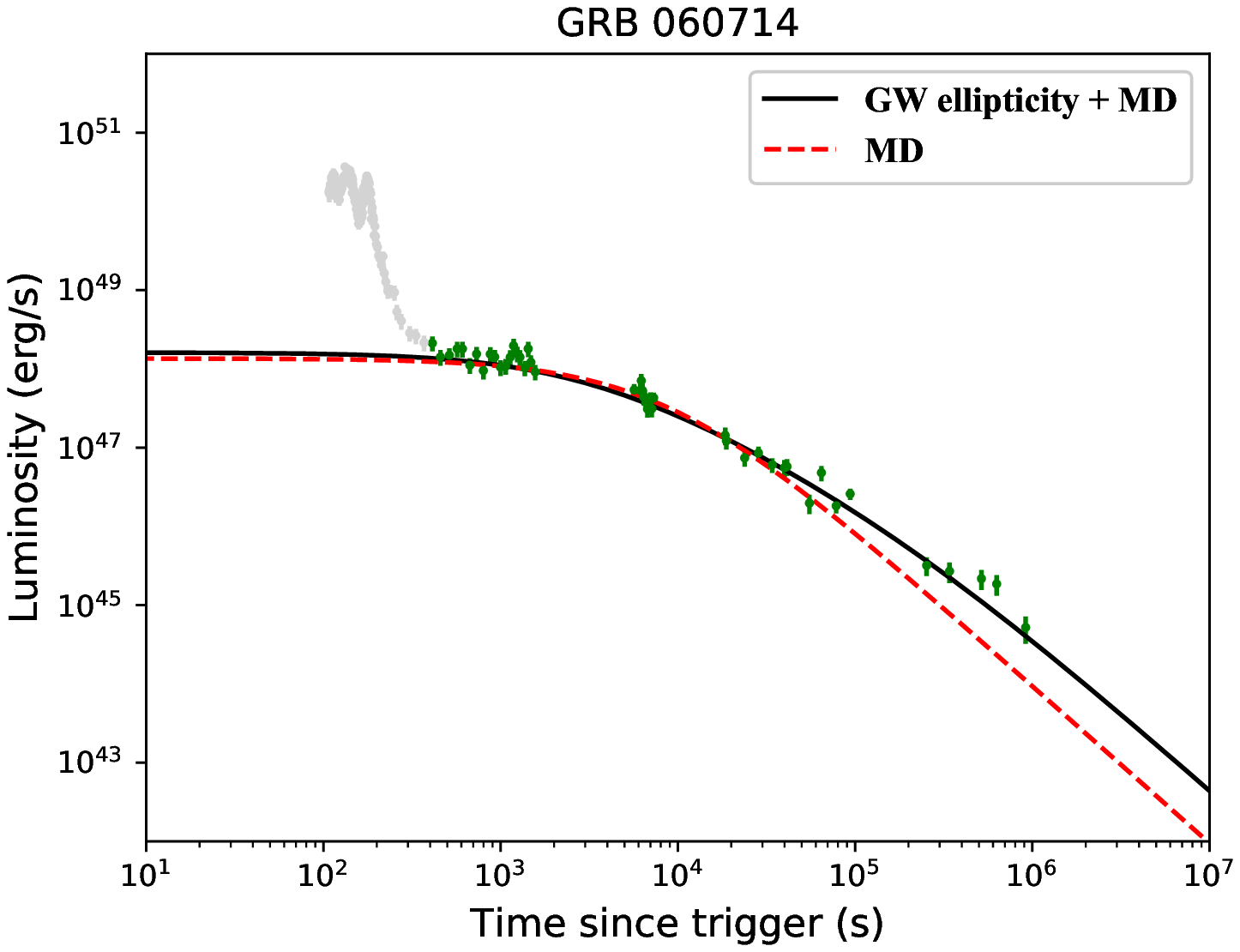}\hspace{-6mm}
\includegraphics[width=0.35\textwidth]{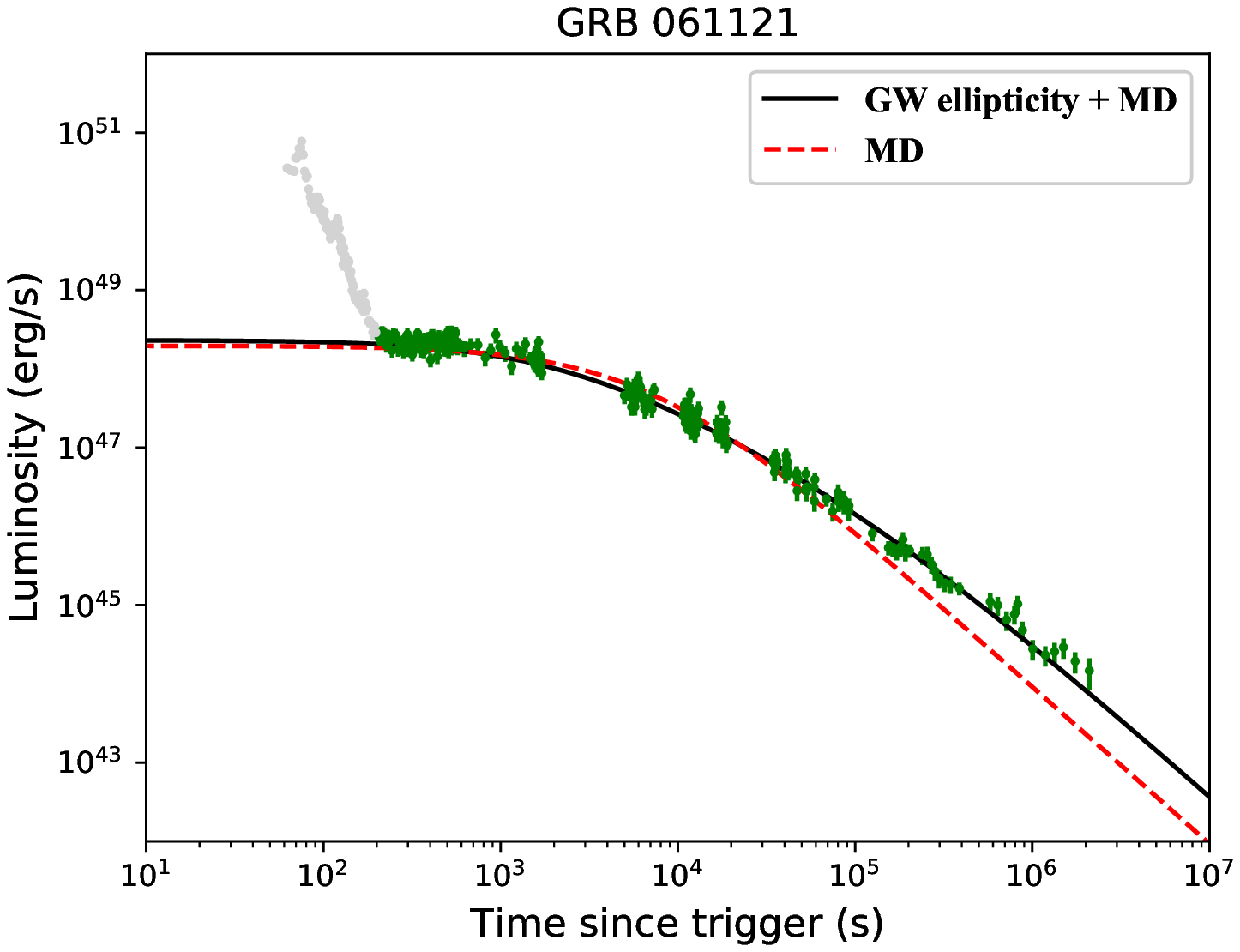}\hspace{-6mm}
\includegraphics[width=0.35\textwidth]{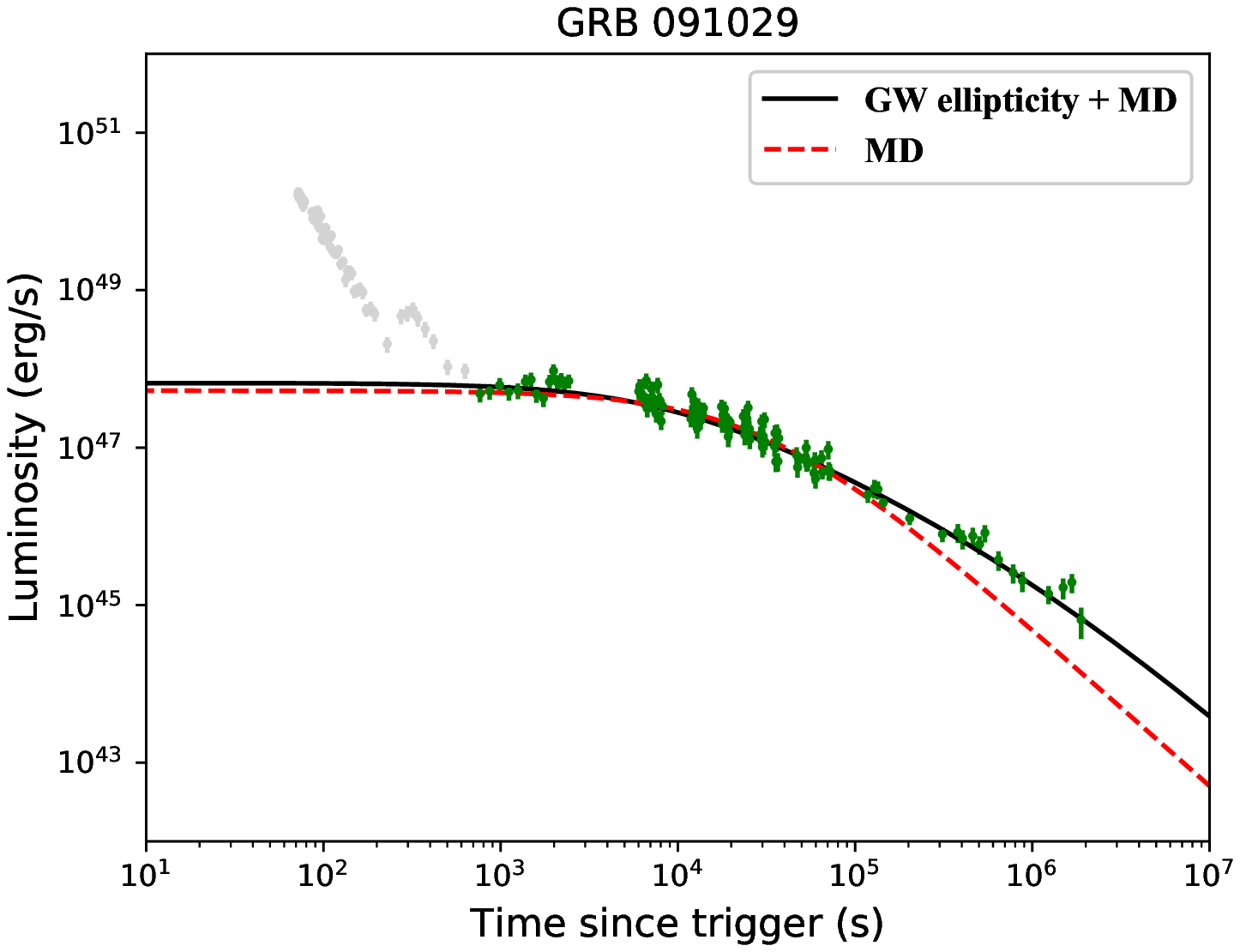}\hspace{-6mm}
\includegraphics[width=0.35\textwidth]{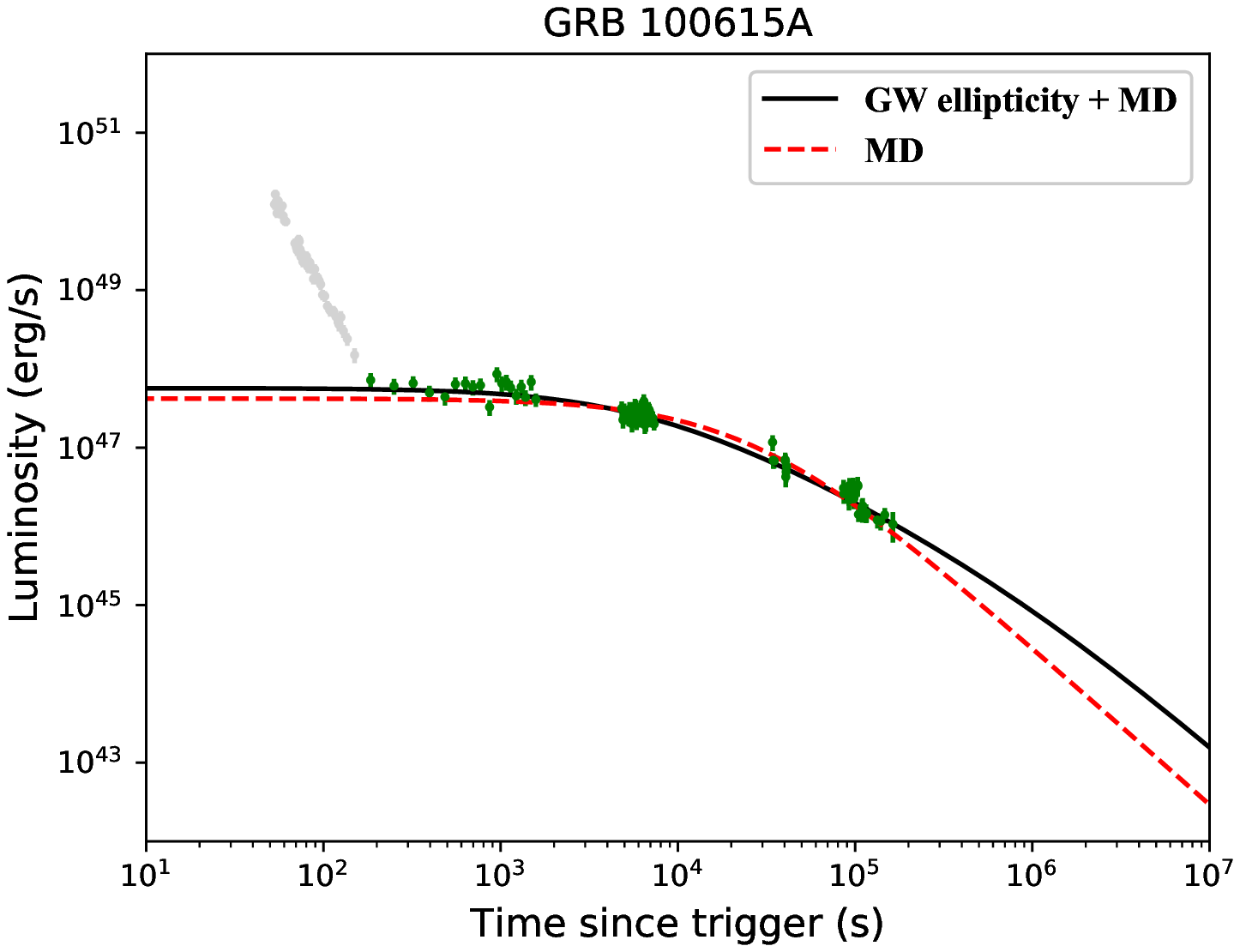}\hspace{-6mm}
\includegraphics[width=0.35\textwidth]{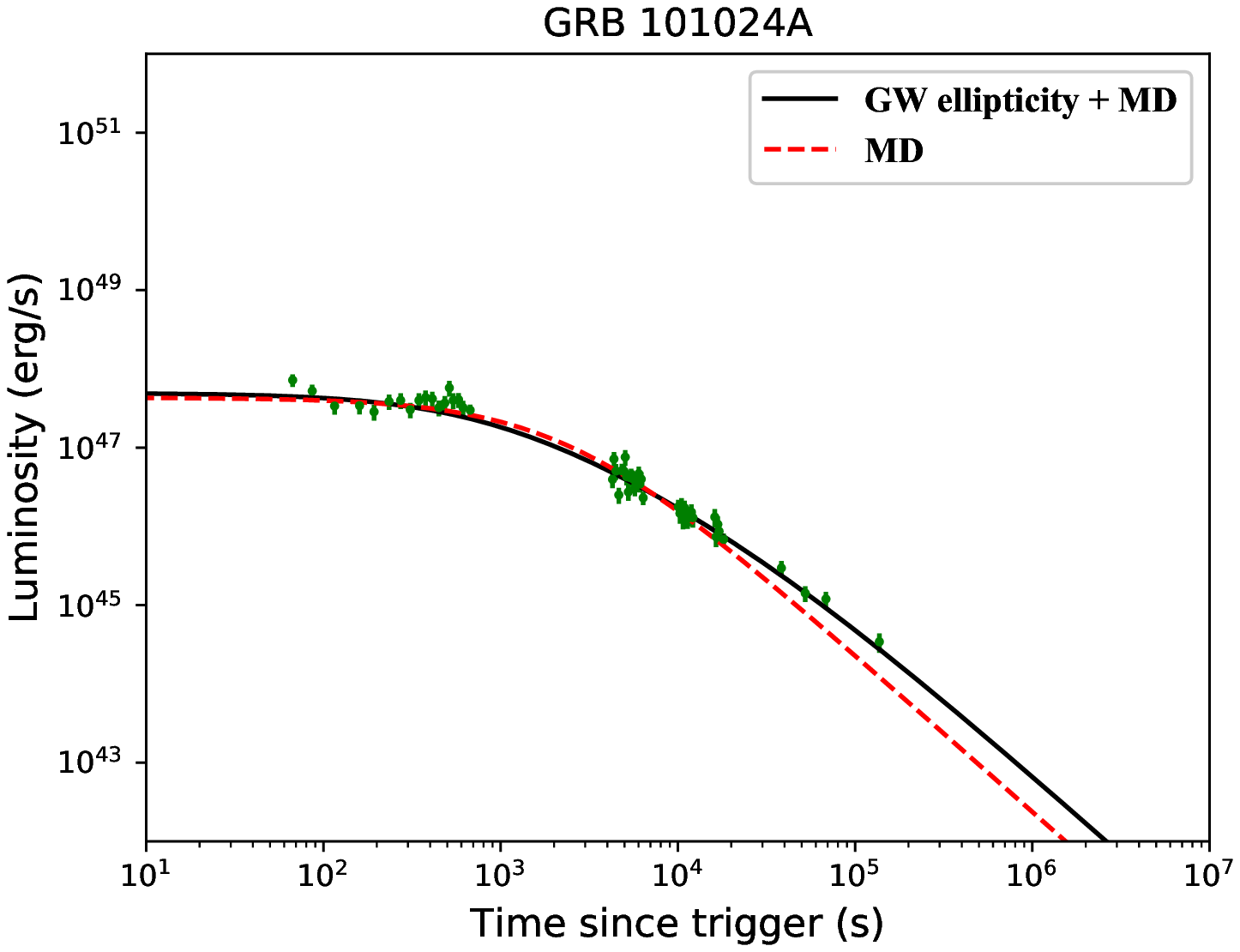}\hspace{-6mm}
\includegraphics[width=0.35\textwidth]{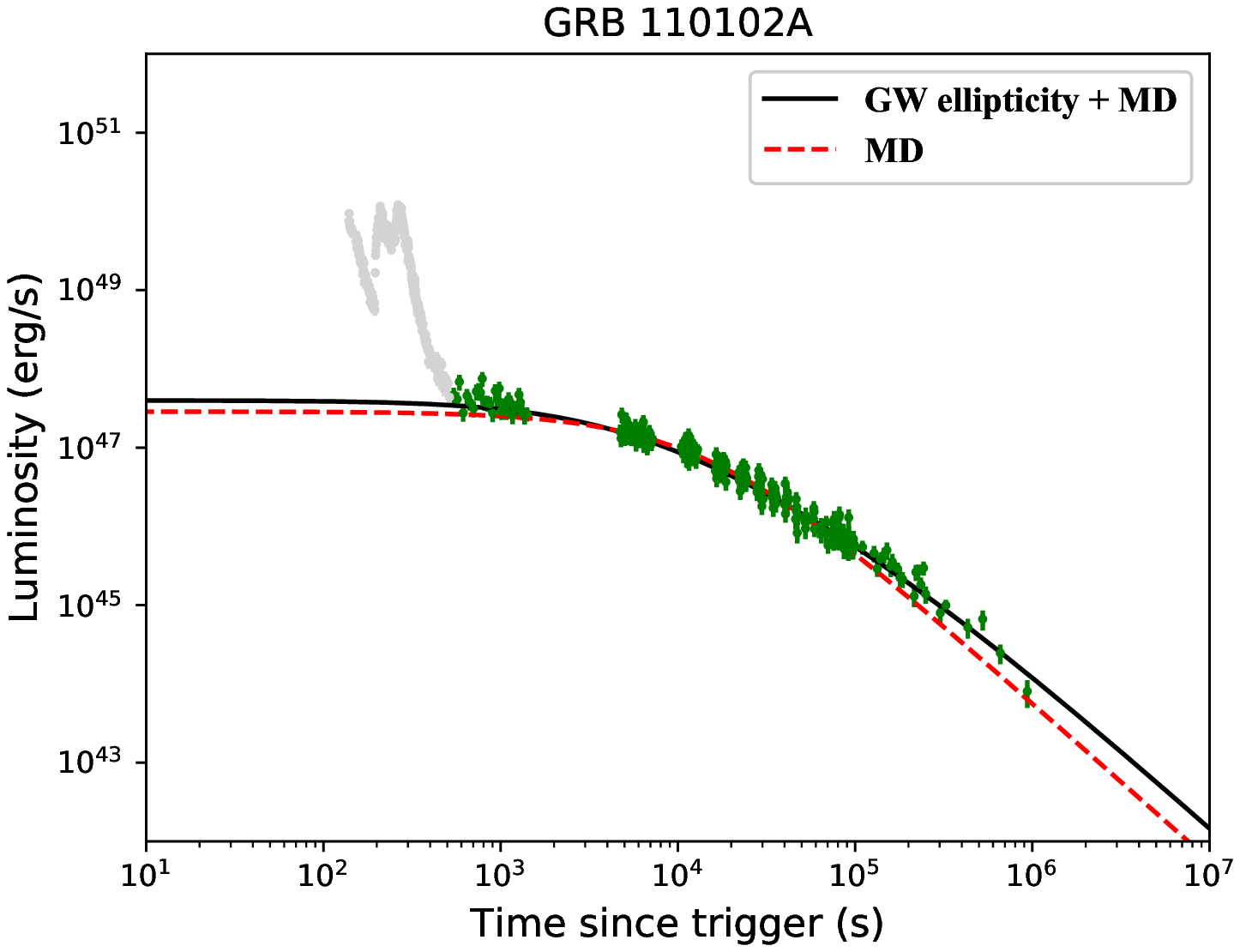}\hspace{-6mm}
\includegraphics[width=0.35\textwidth]{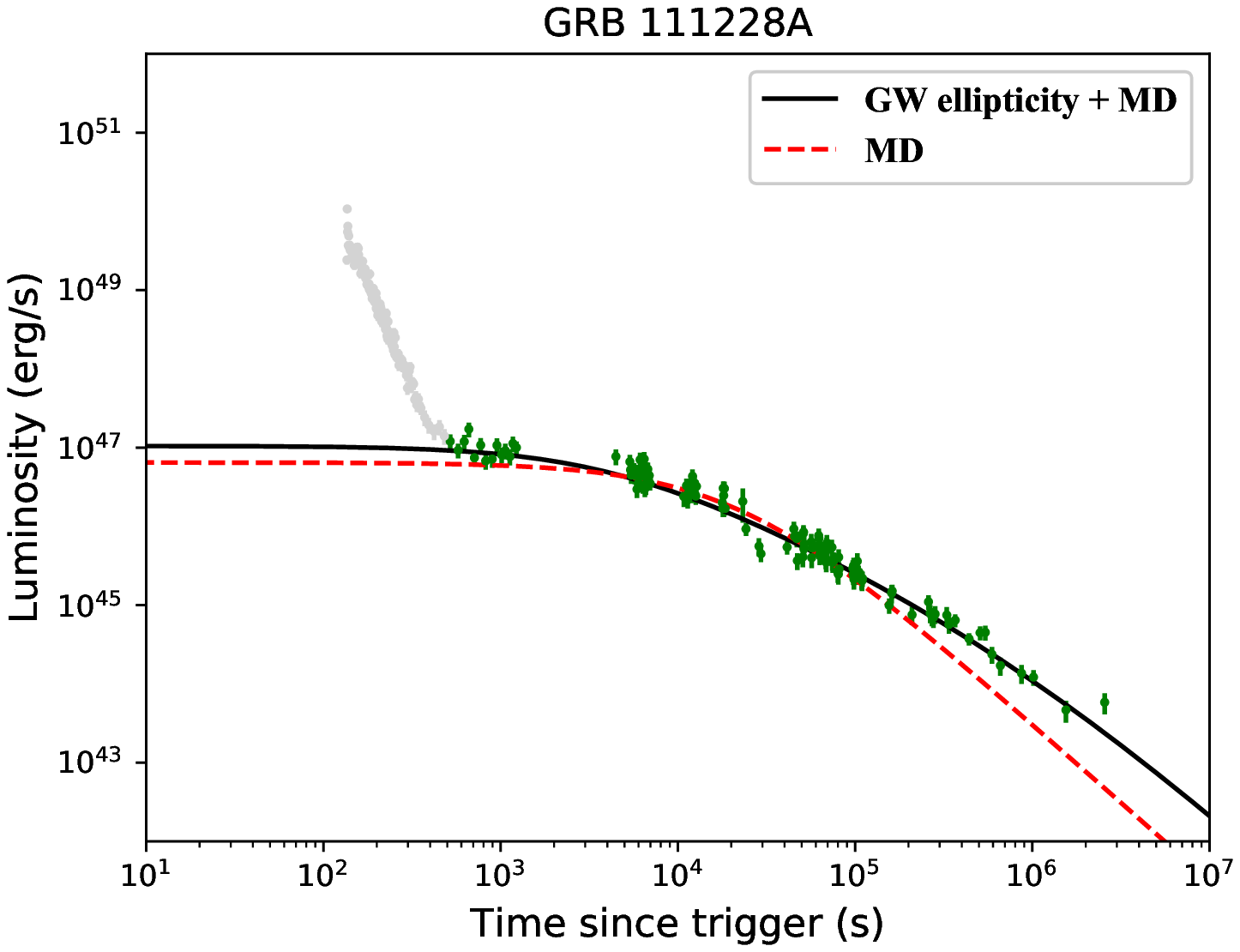}\hspace{-6mm}
\includegraphics[width=0.35\textwidth]{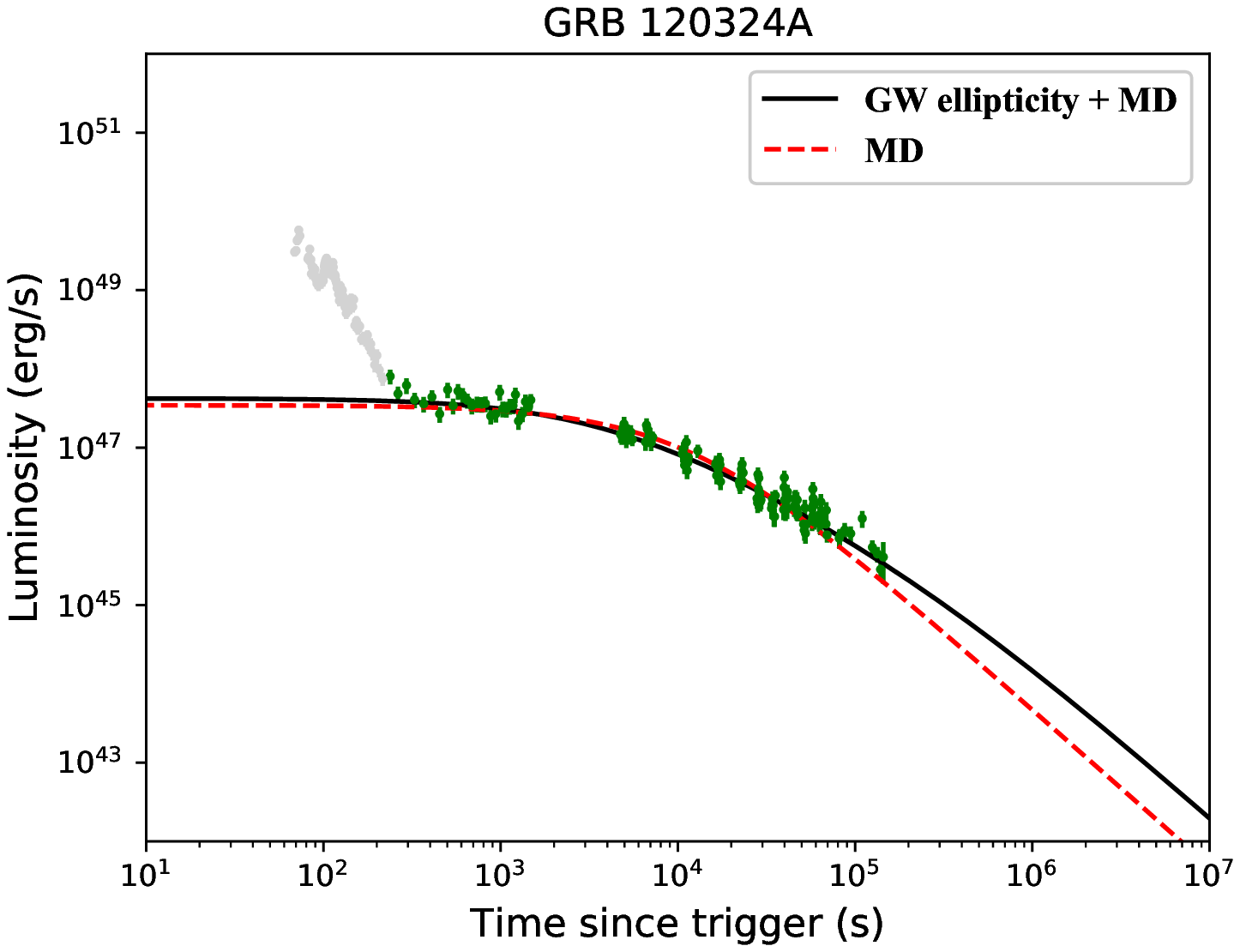}\hspace{-6mm}
\includegraphics[width=0.35\textwidth]{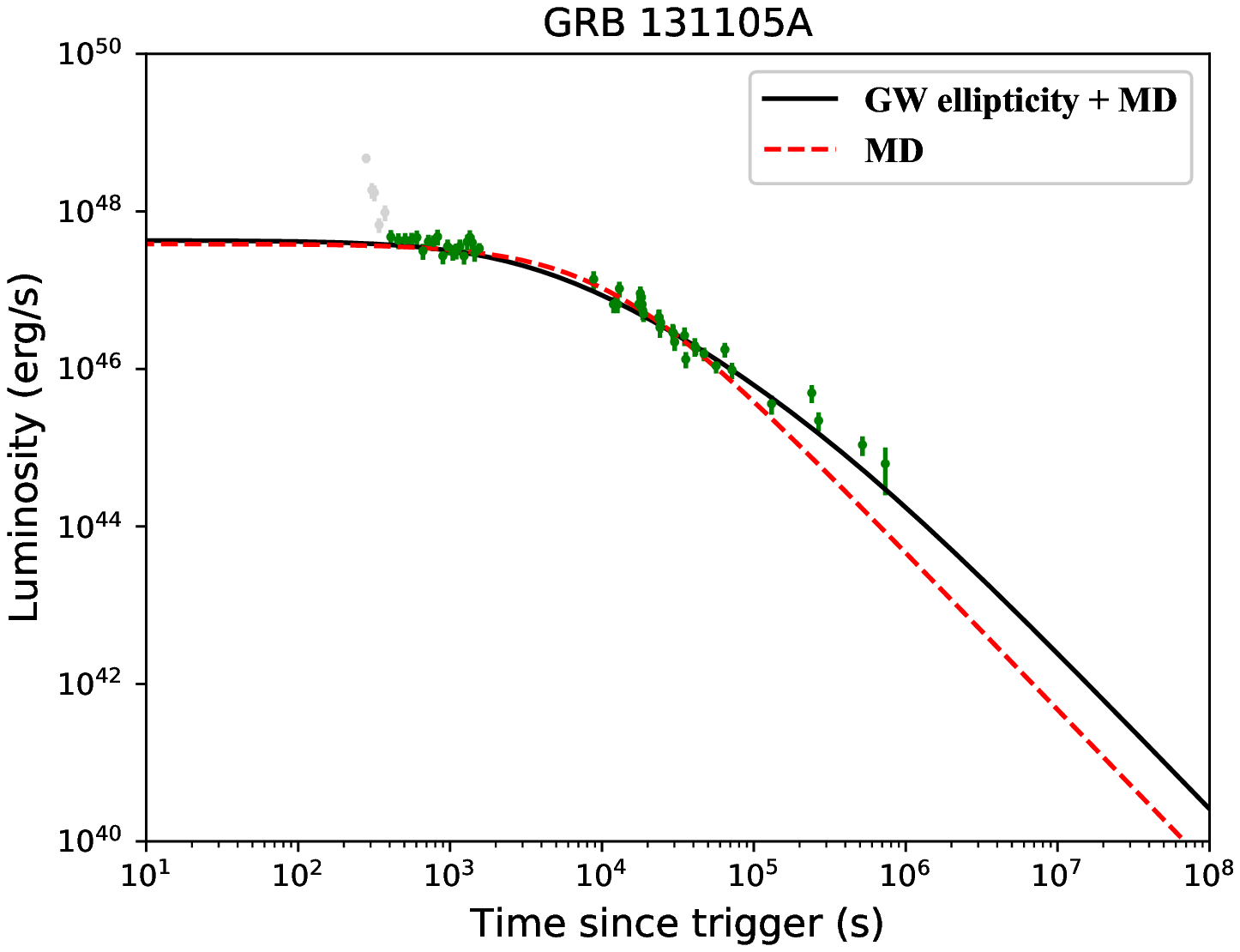}\hspace{-6mm}
\includegraphics[width=0.35\textwidth]{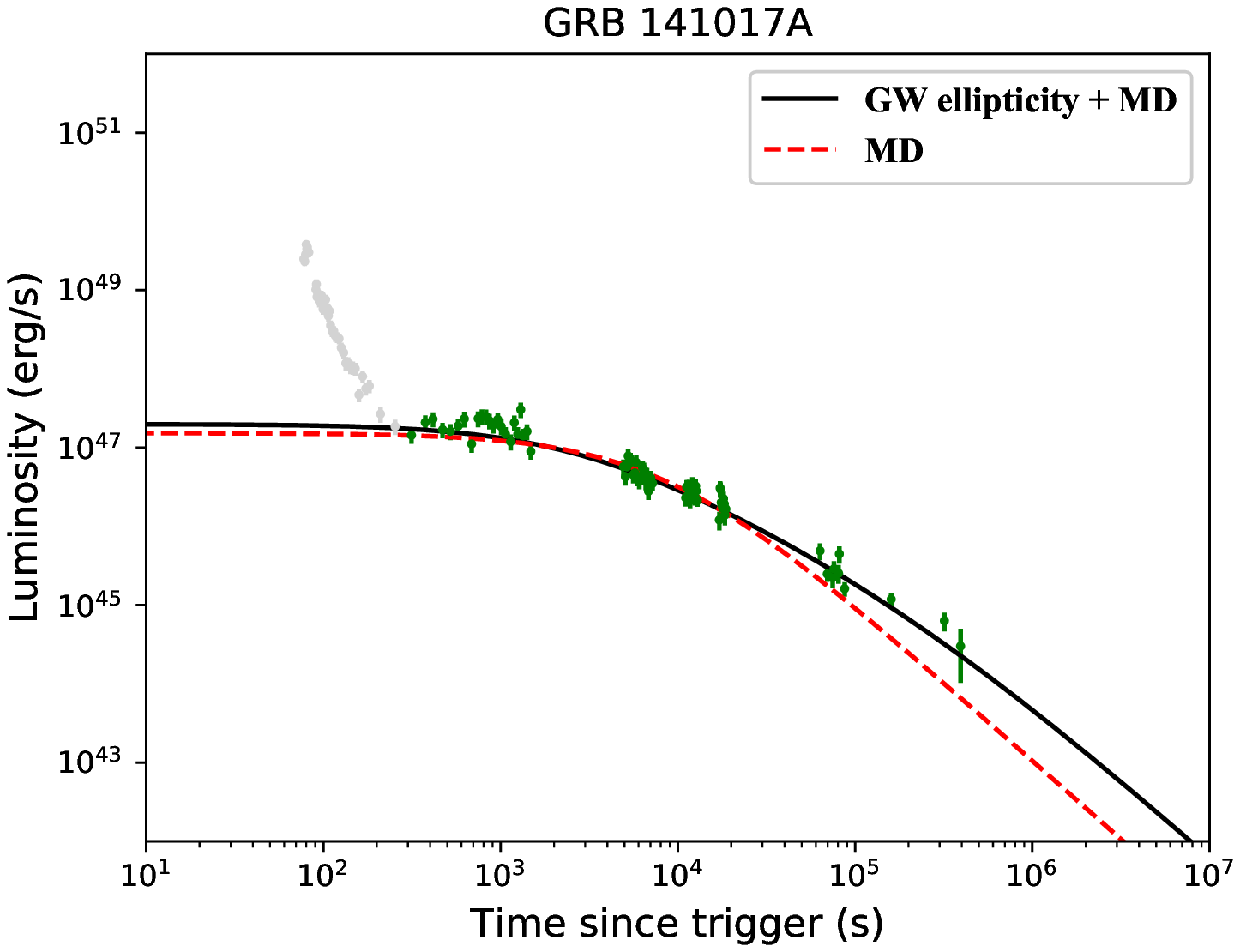}\hspace{-6mm}
\includegraphics[width=0.35\textwidth]{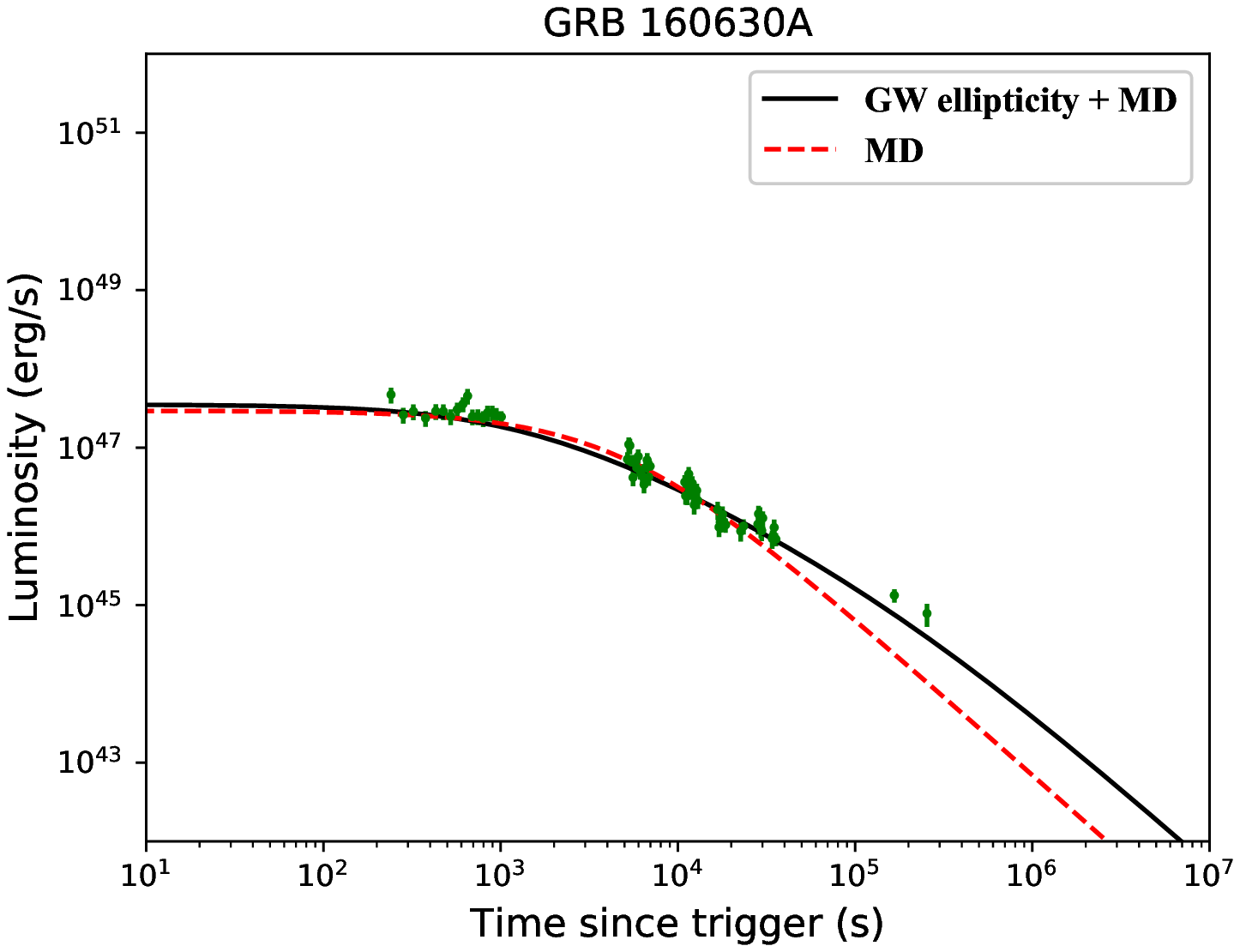}\hspace{-6mm}
\includegraphics[width=0.35\textwidth]{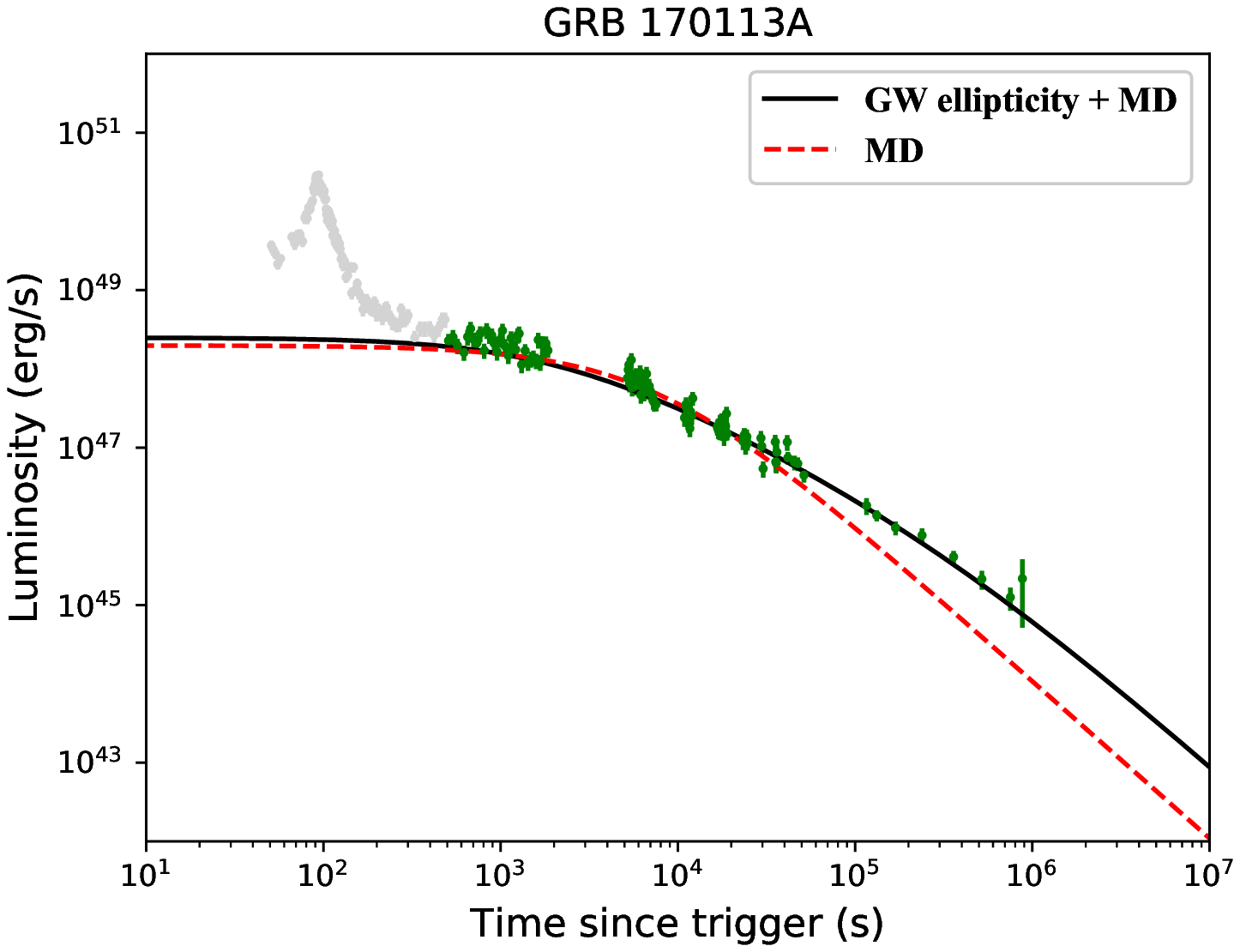}\hspace{-6mm}
\includegraphics[width=0.35\textwidth]{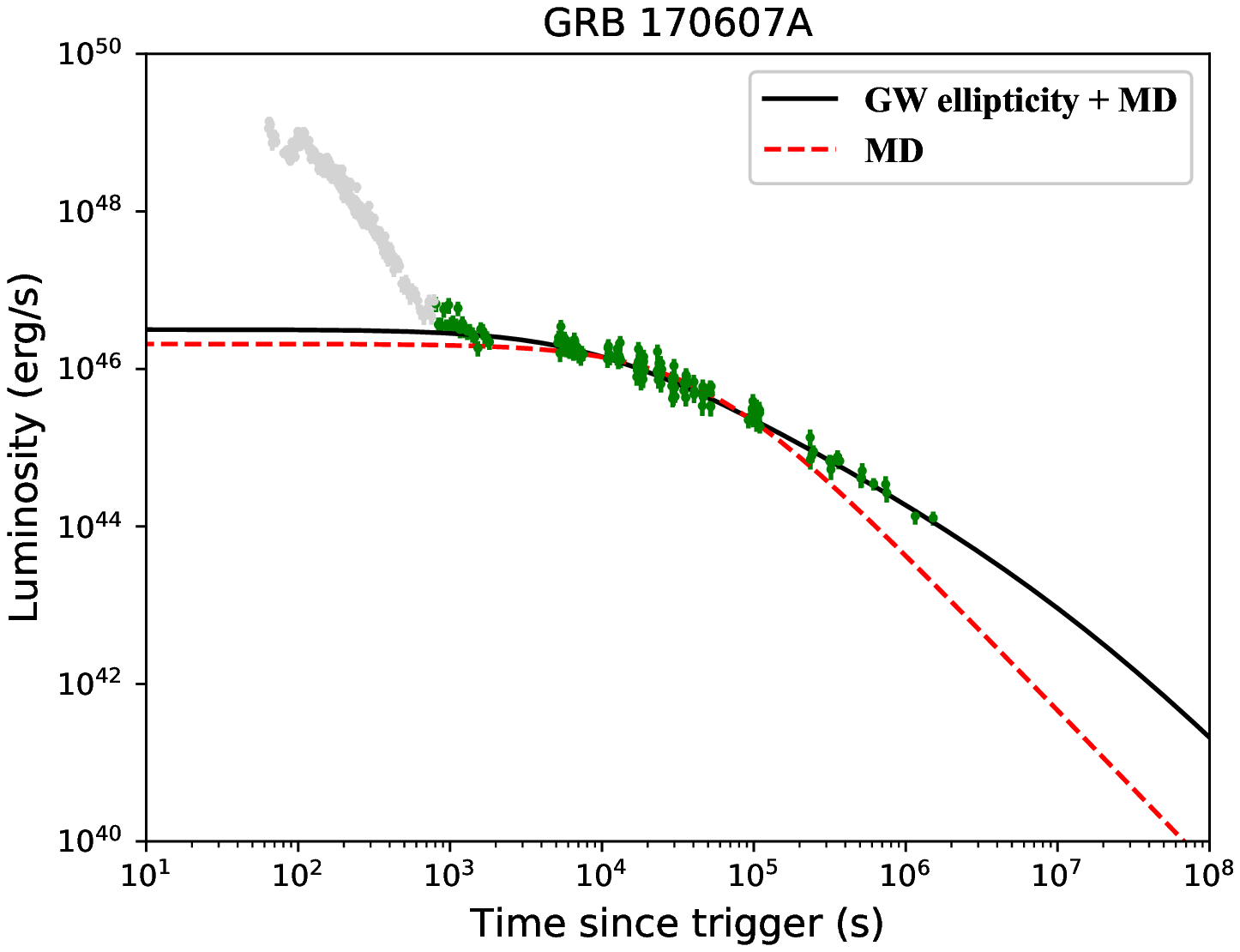}\hspace{-6mm}
\includegraphics[width=0.35\textwidth]{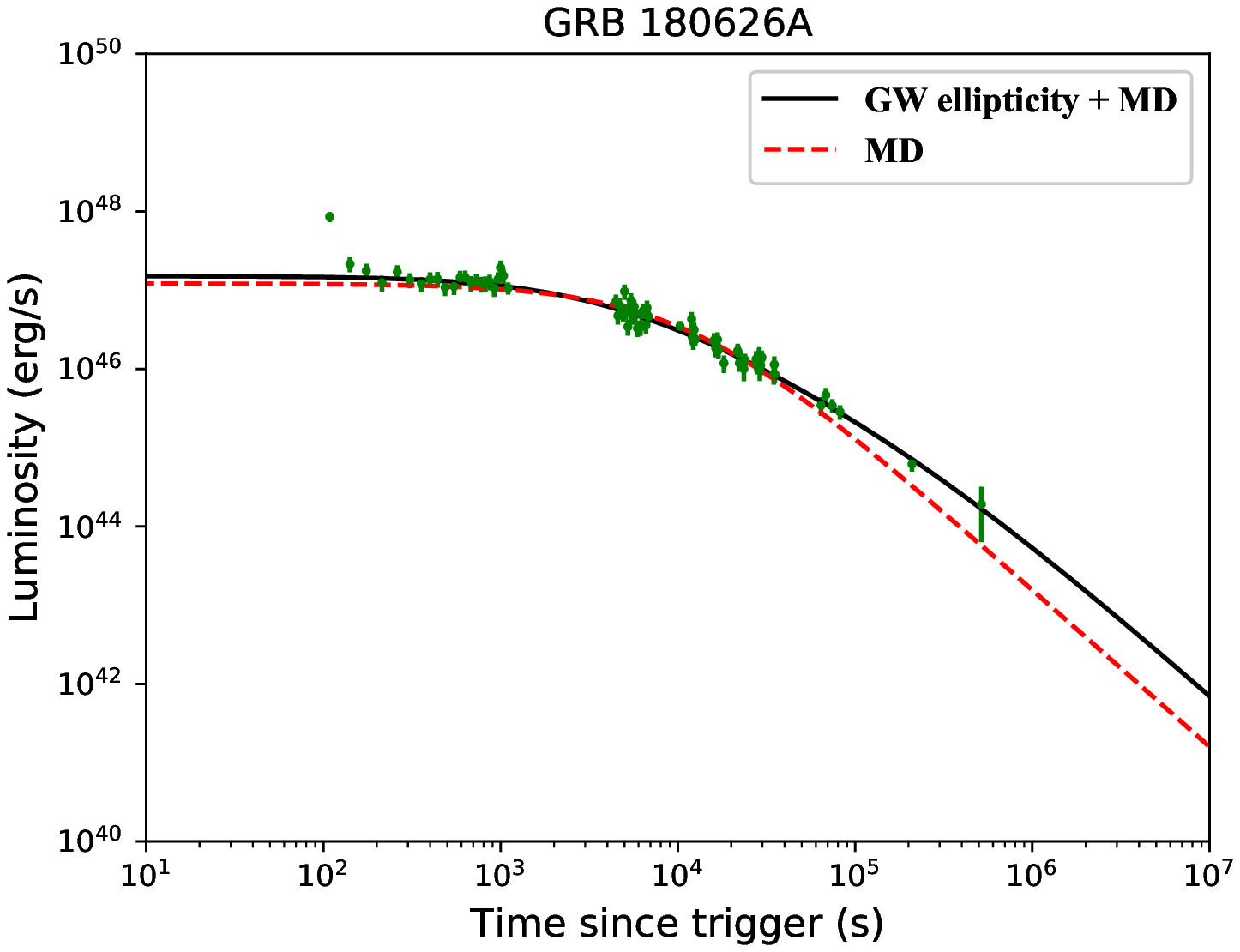}\hspace{-6mm}
\includegraphics[width=0.35\textwidth]{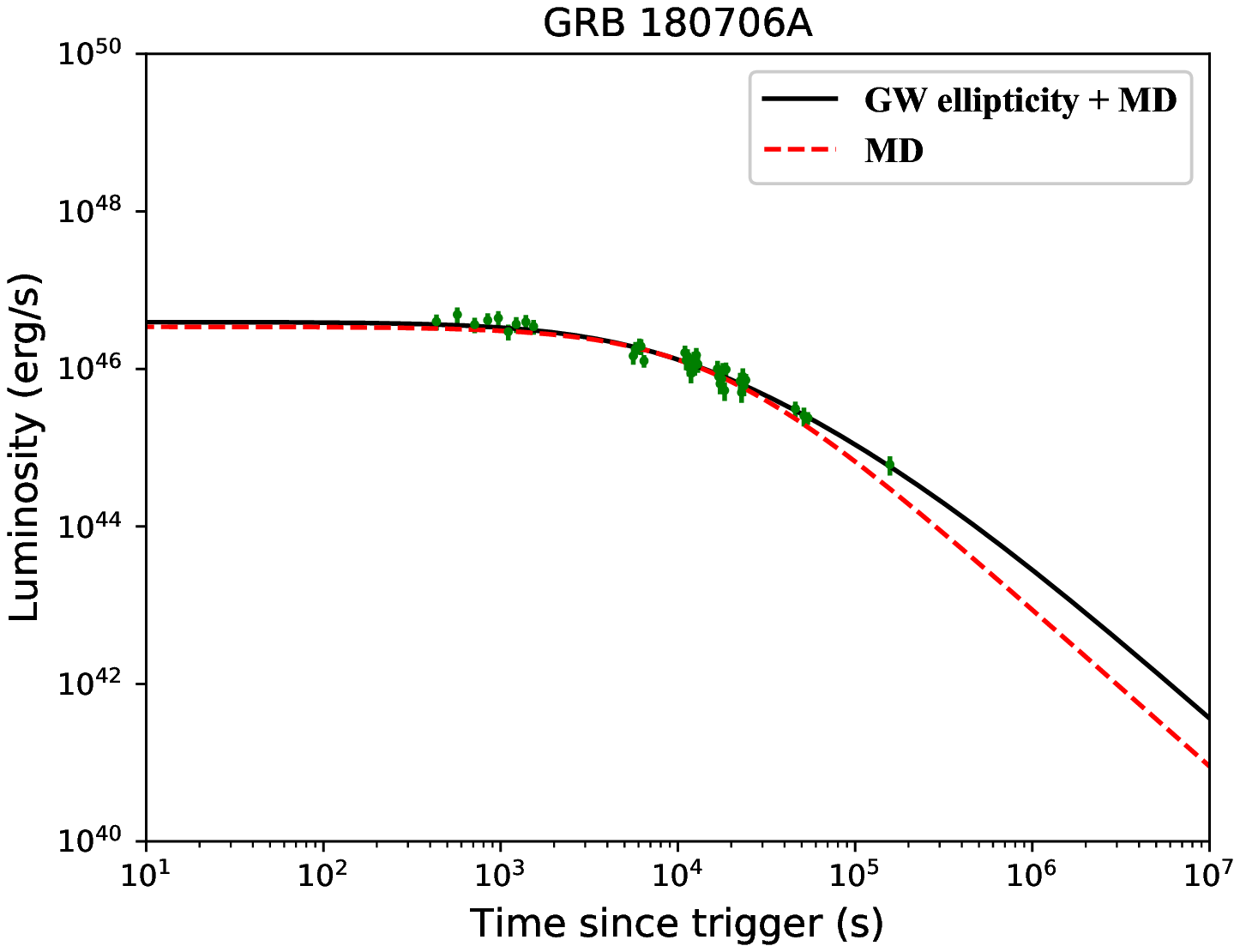}\hspace{-6mm}

\caption {15 LGRB XRT light curves which can be well fitted with the hybrid model. The green data points are the XRT light curves of LGRBs. The black curves (red dashed curves) show the best-fitting results for the hybrid model (MD model). }
\label{fig-GWsamplexrt}
\end{figure*}

\begin{figure*}
\centering
\includegraphics[width=0.35\textwidth]{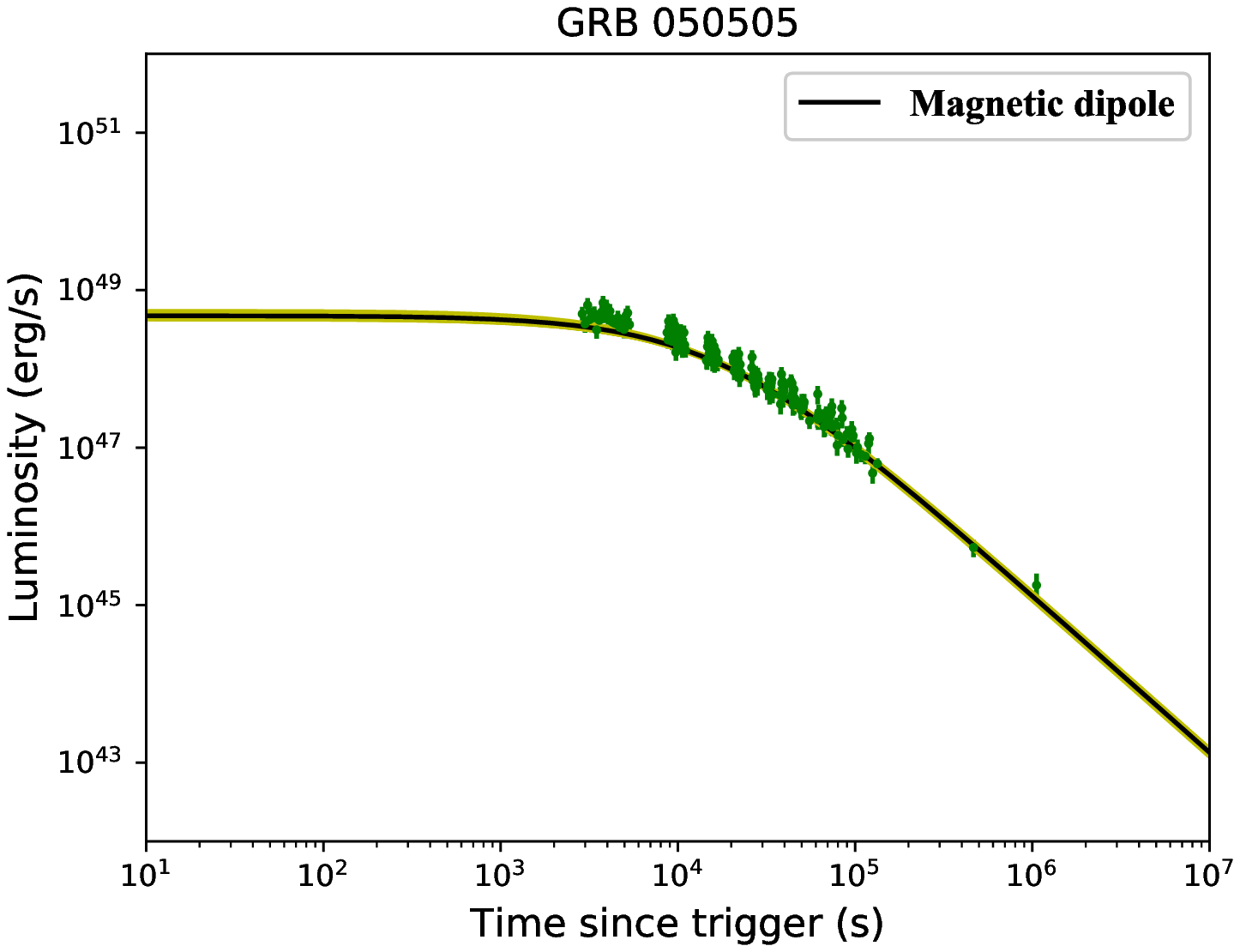}\hspace{-6mm}
\includegraphics[width=0.35\textwidth]{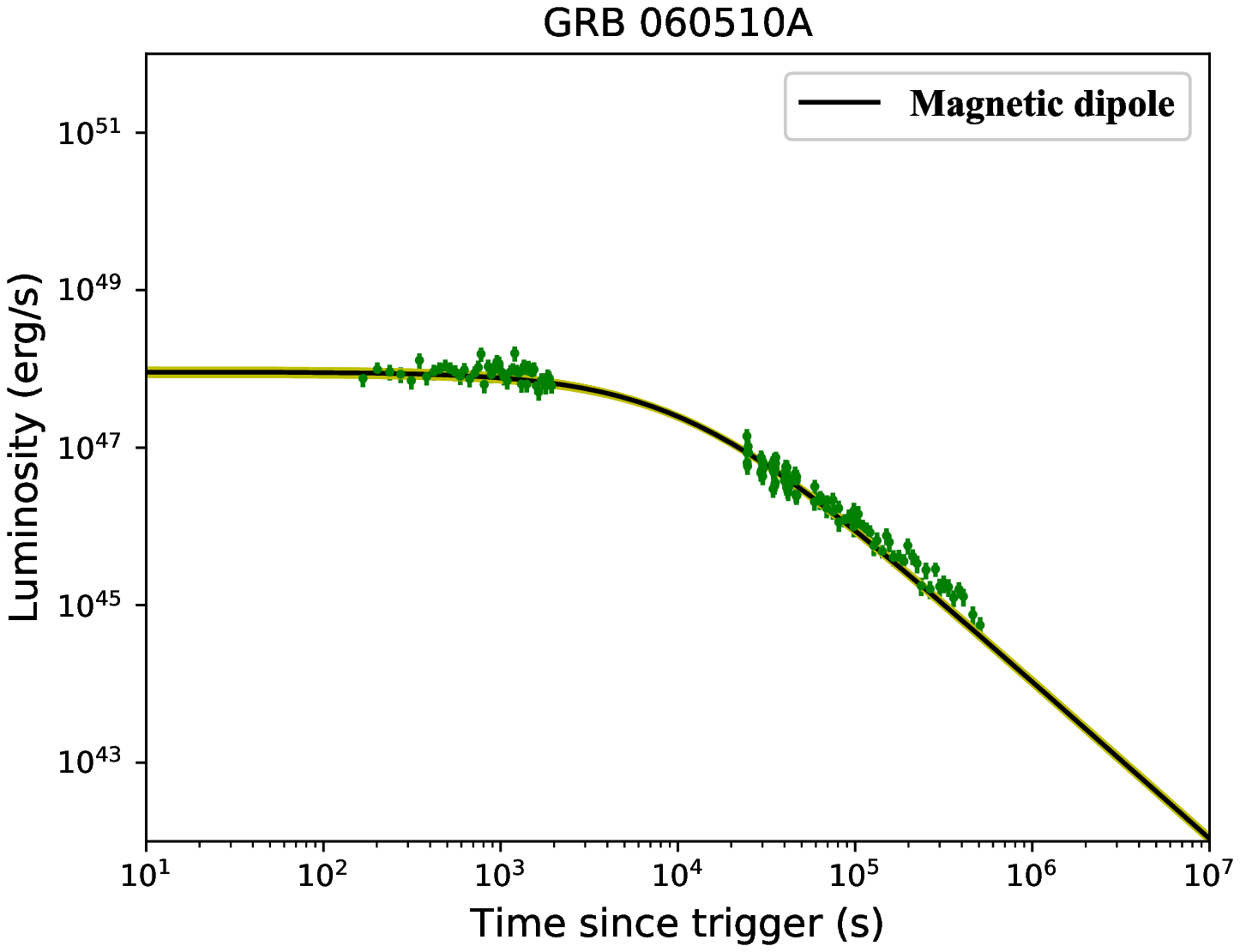}\hspace{-6mm}
\includegraphics[width=0.35\textwidth]{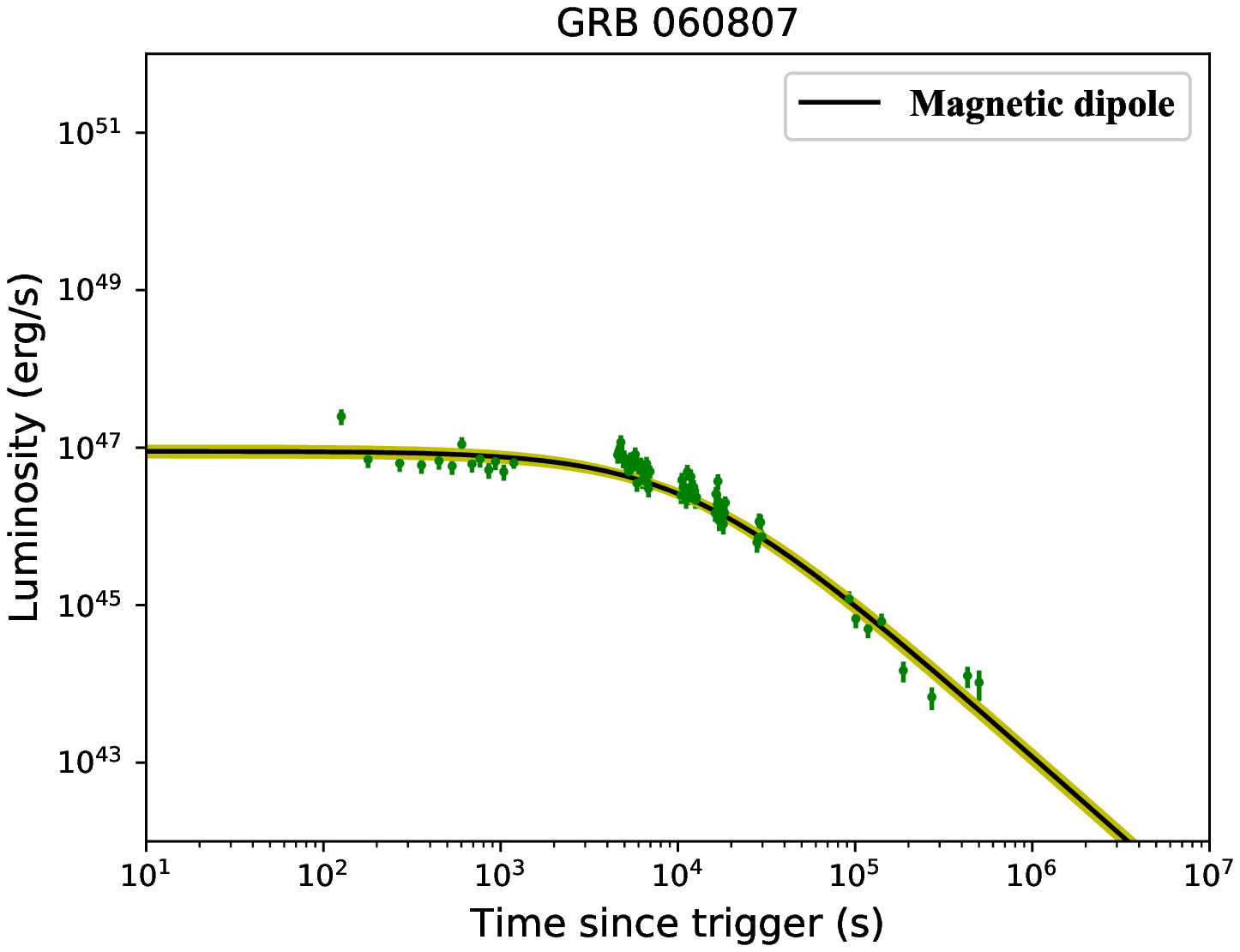}\hspace{-6mm}
\includegraphics[width=0.35\textwidth]{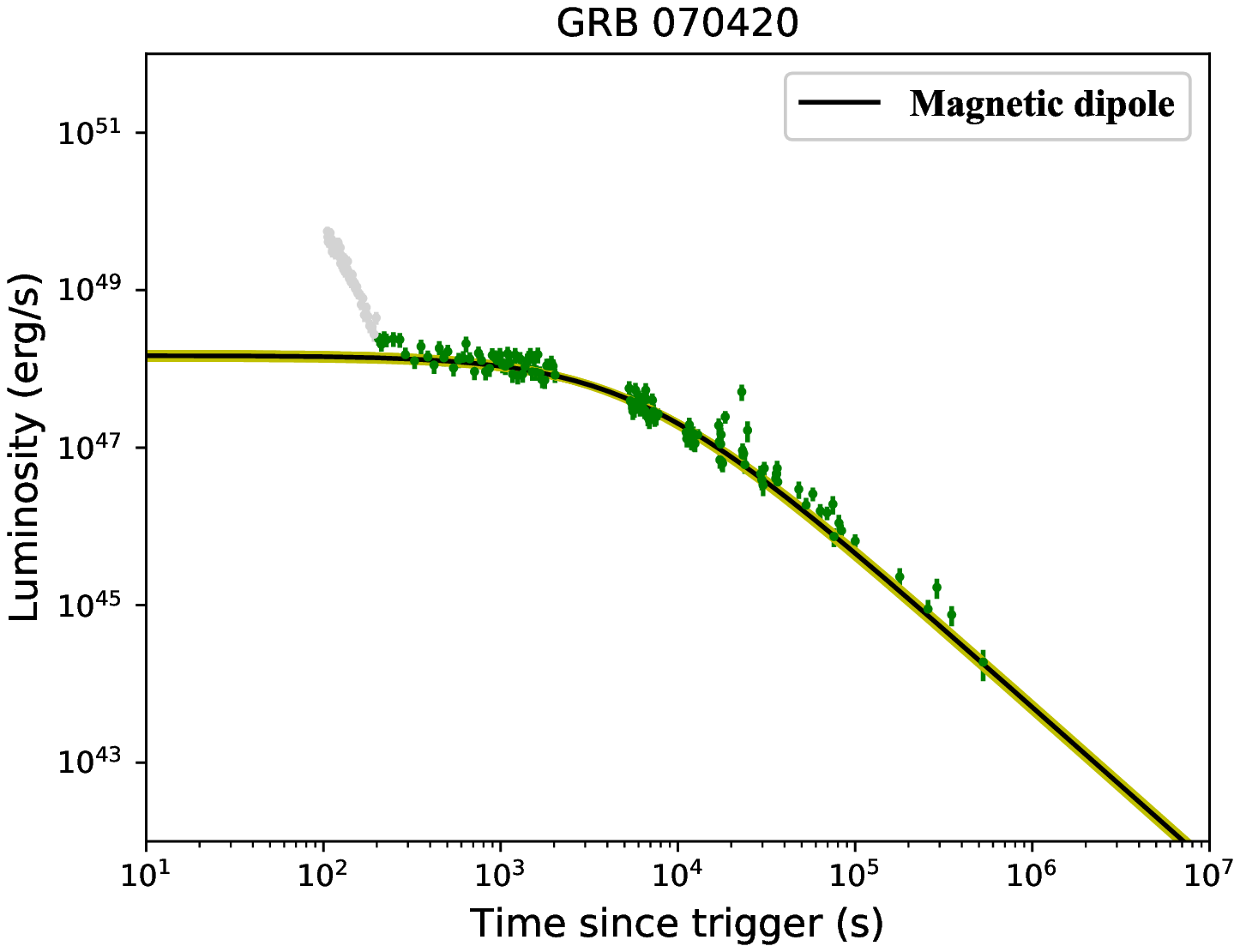}\hspace{-6mm}
\includegraphics[width=0.35\textwidth]{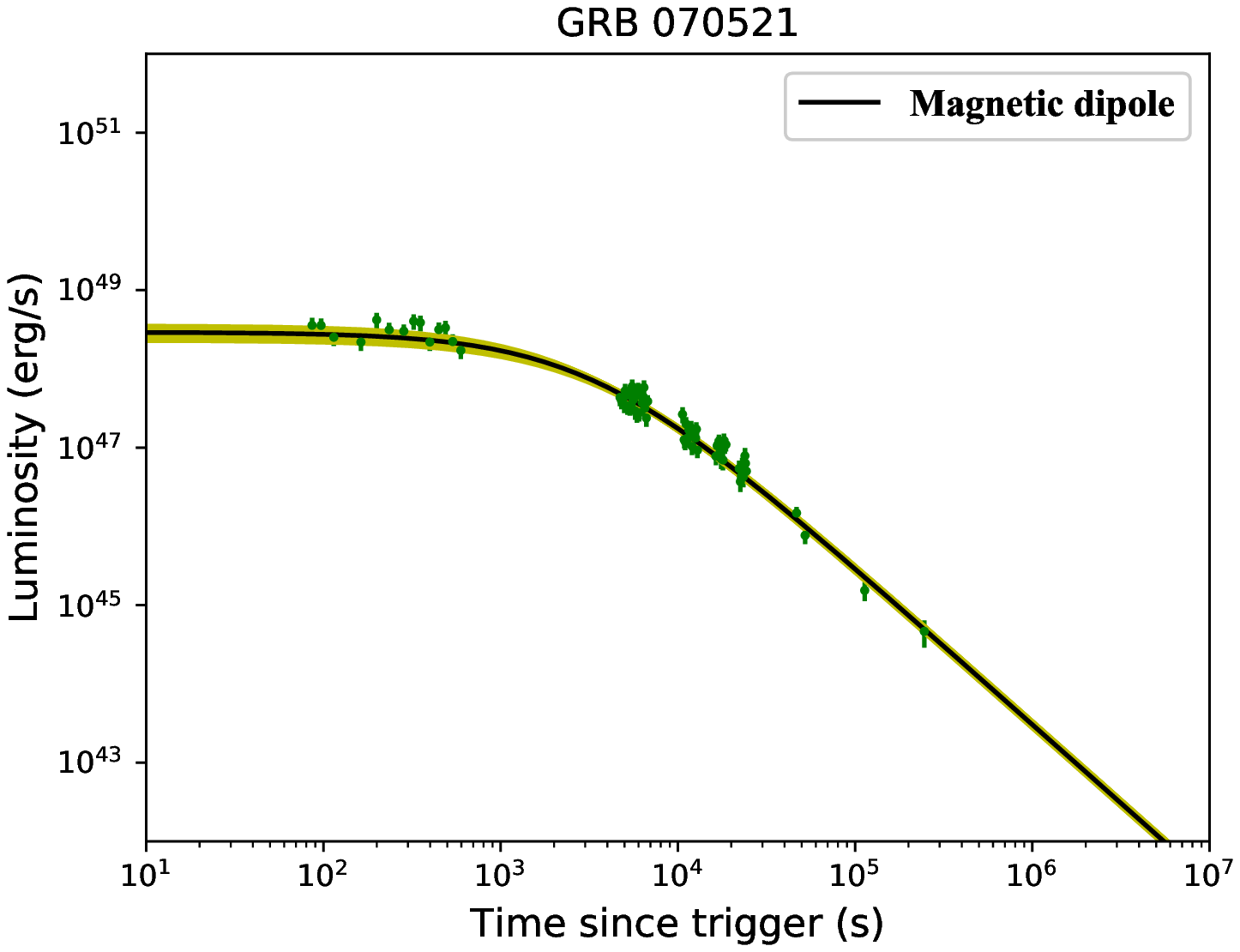}\hspace{-6mm}
\includegraphics[width=0.35\textwidth]{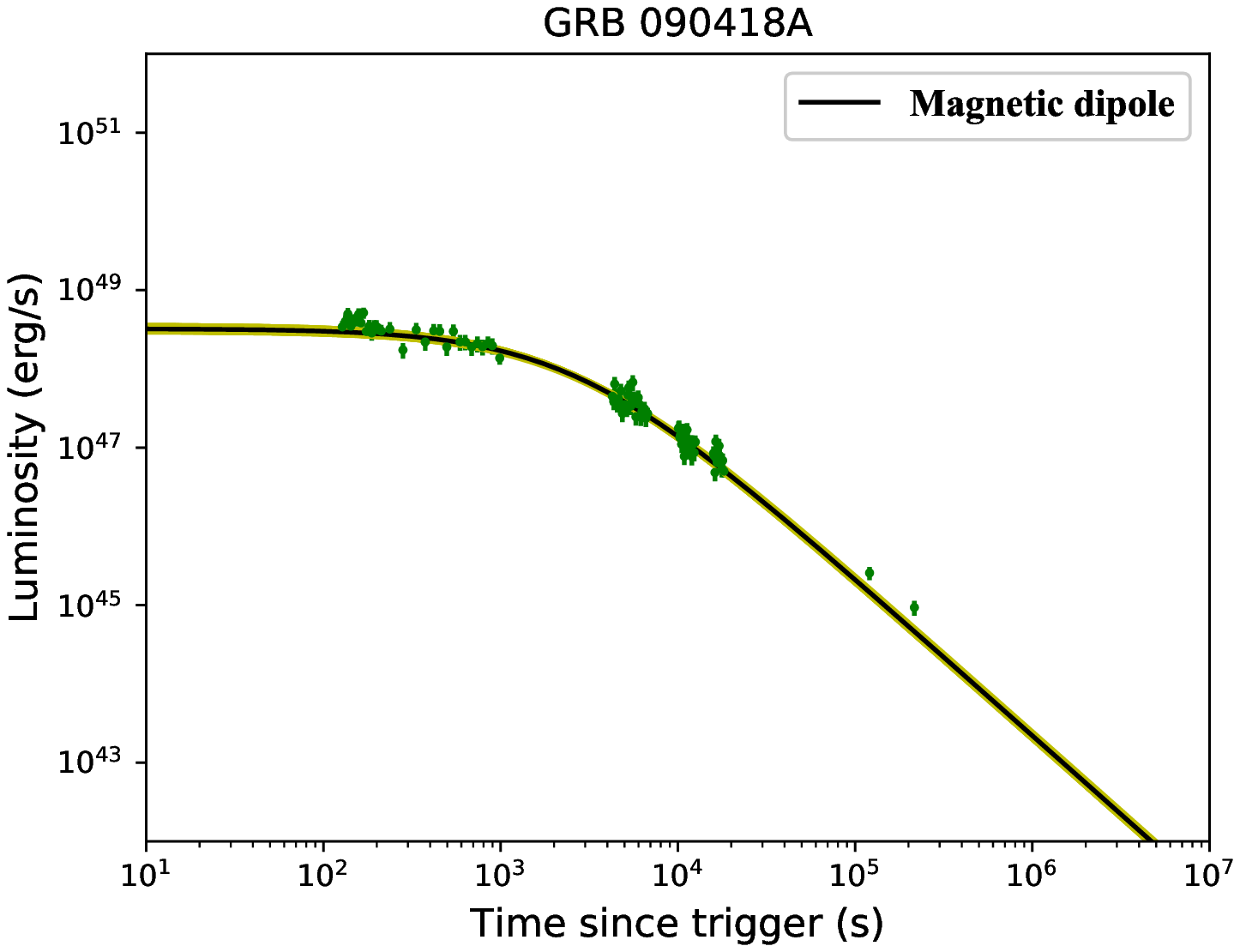}\hspace{-6mm}
\includegraphics[width=0.35\textwidth]{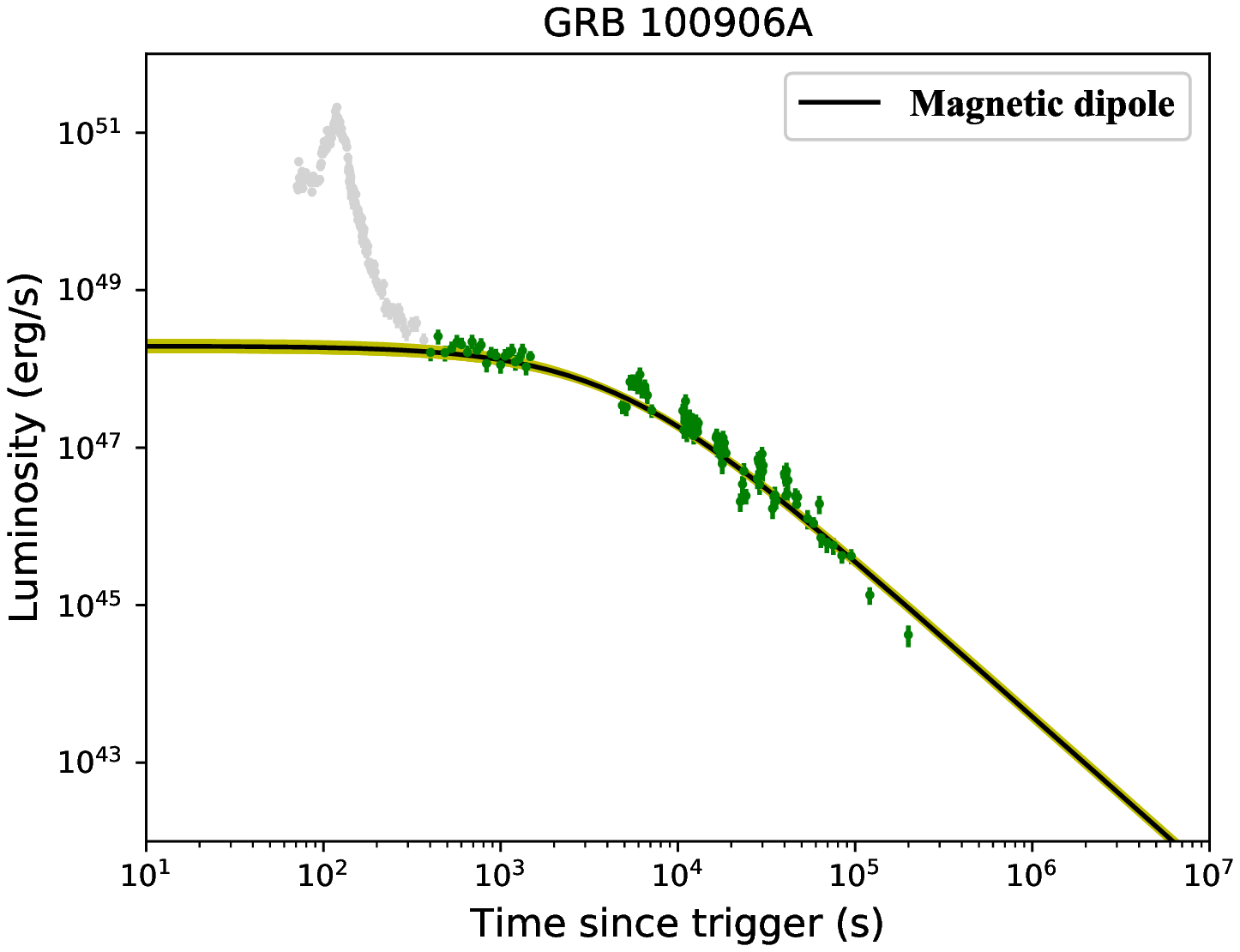}\hspace{-6mm}
\includegraphics[width=0.35\textwidth]{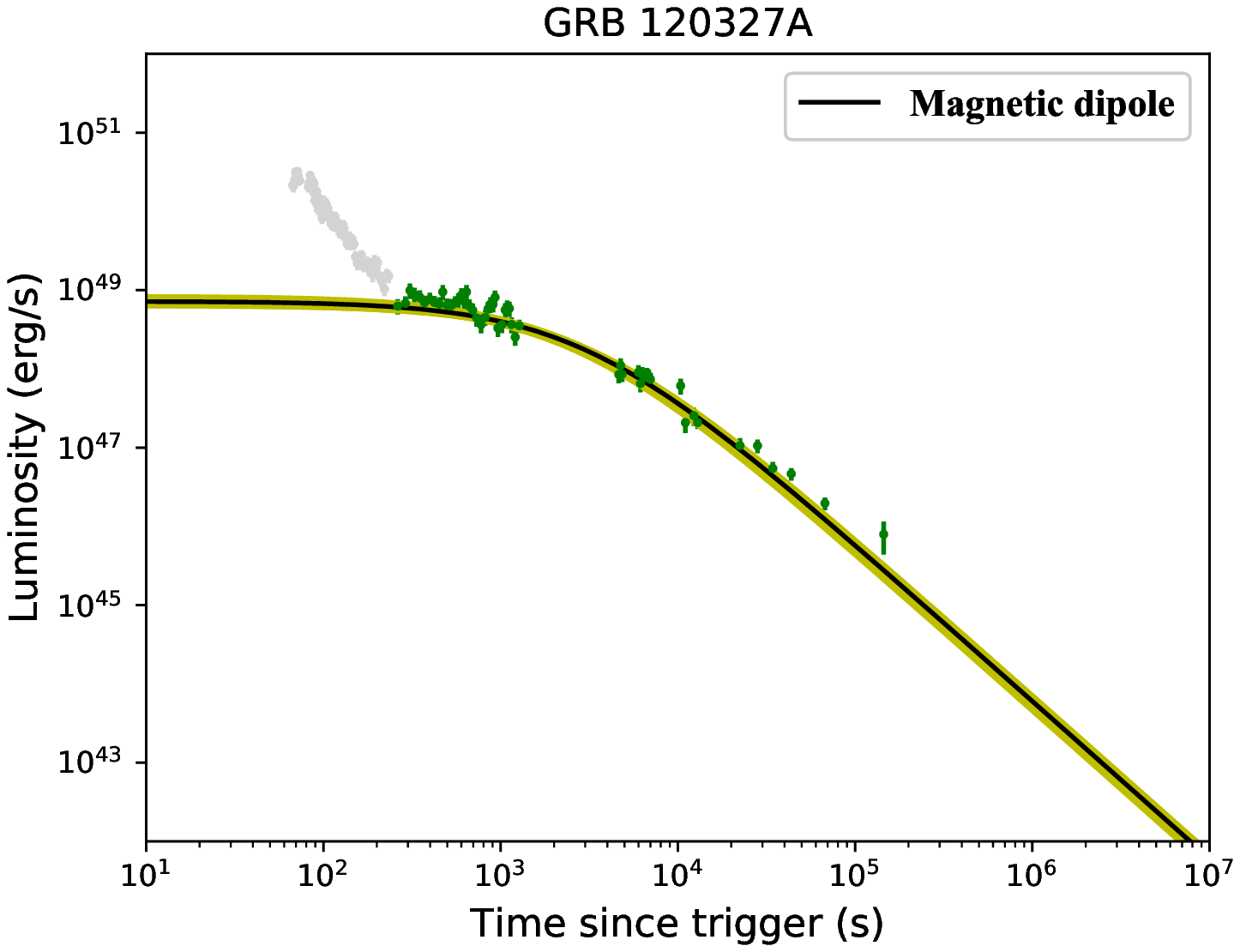}\hspace{-6mm}
\includegraphics[width=0.35\textwidth]{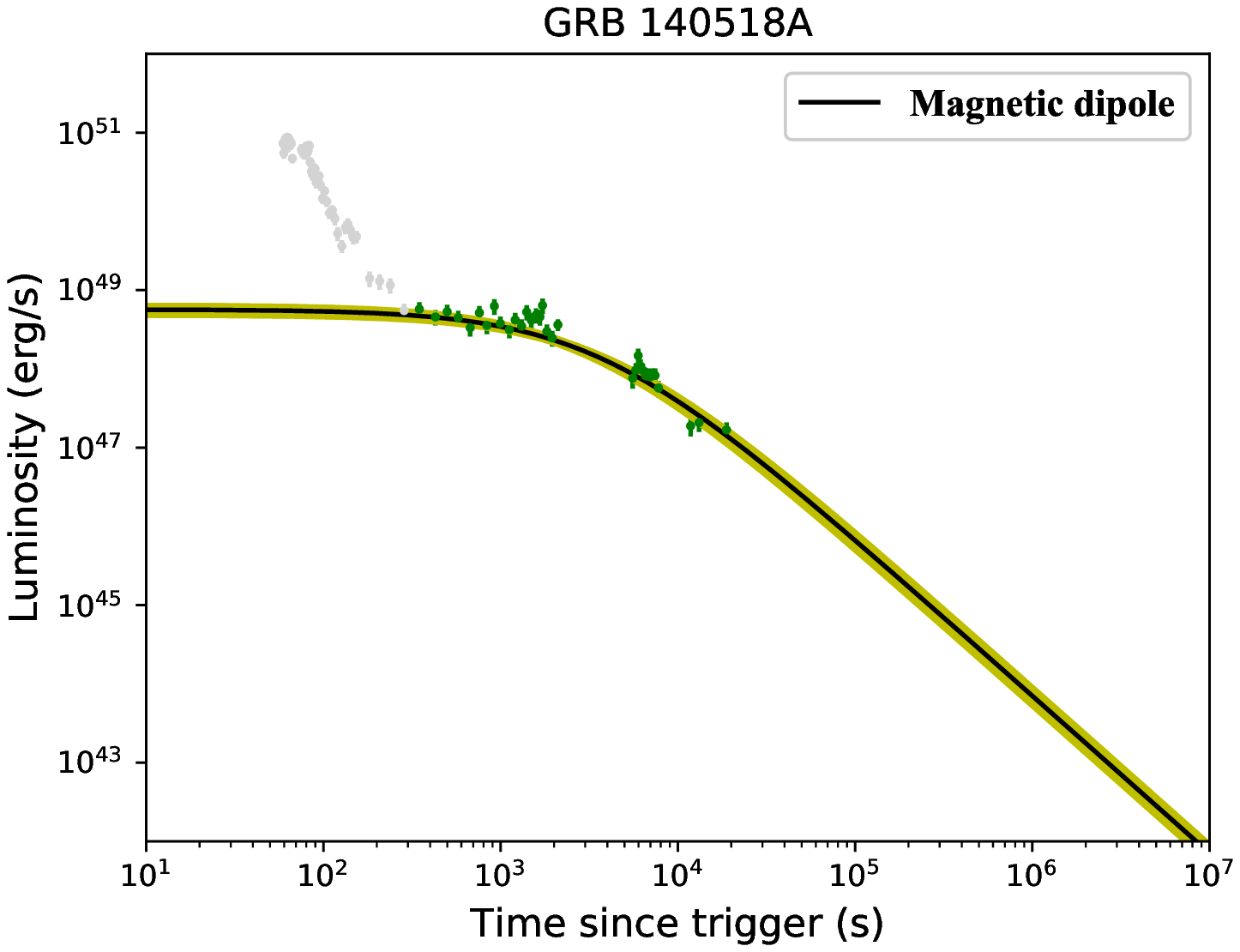}\hspace{-6mm}
\includegraphics[width=0.35\textwidth]{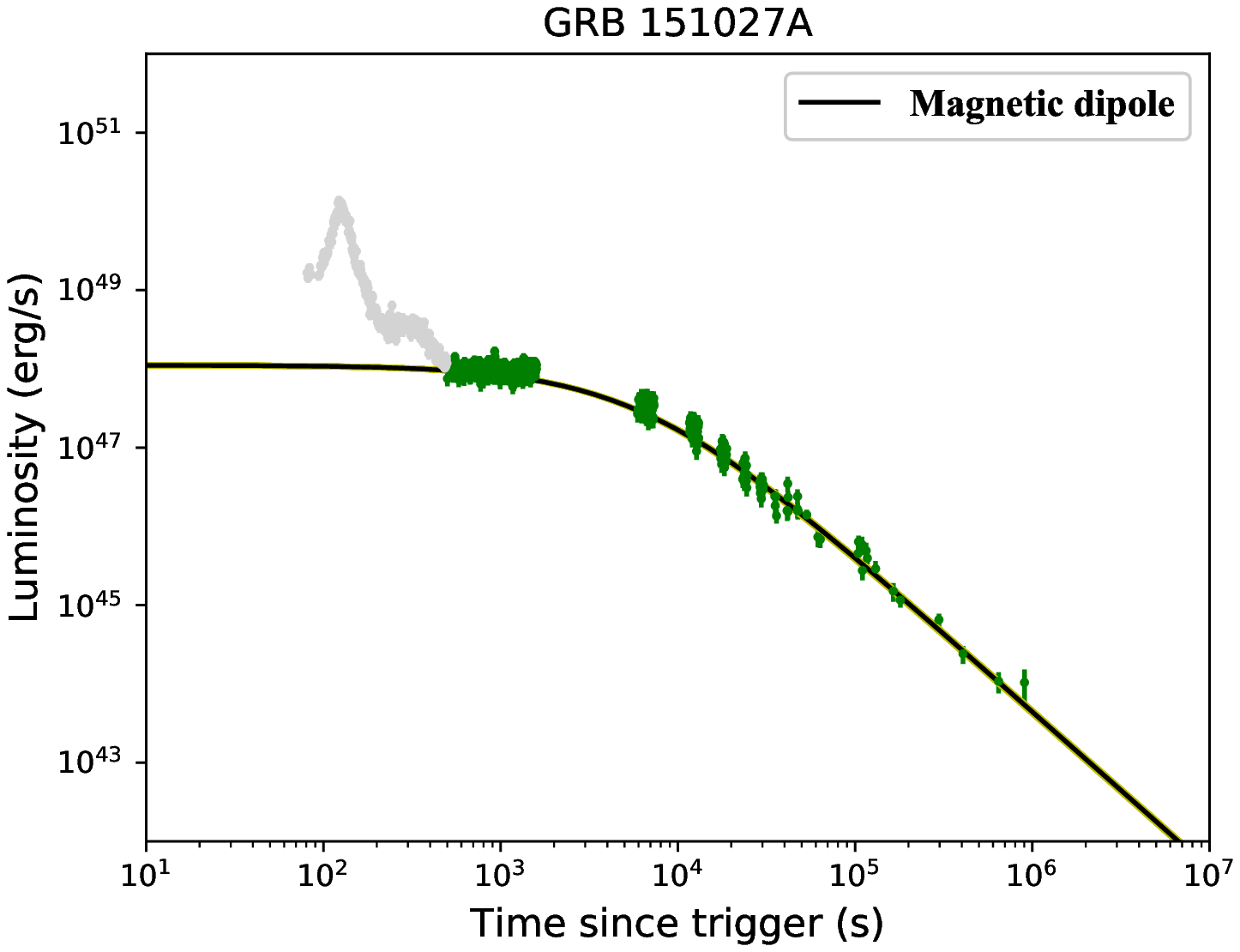}\hspace{-6mm}
\includegraphics[width=0.35\textwidth]{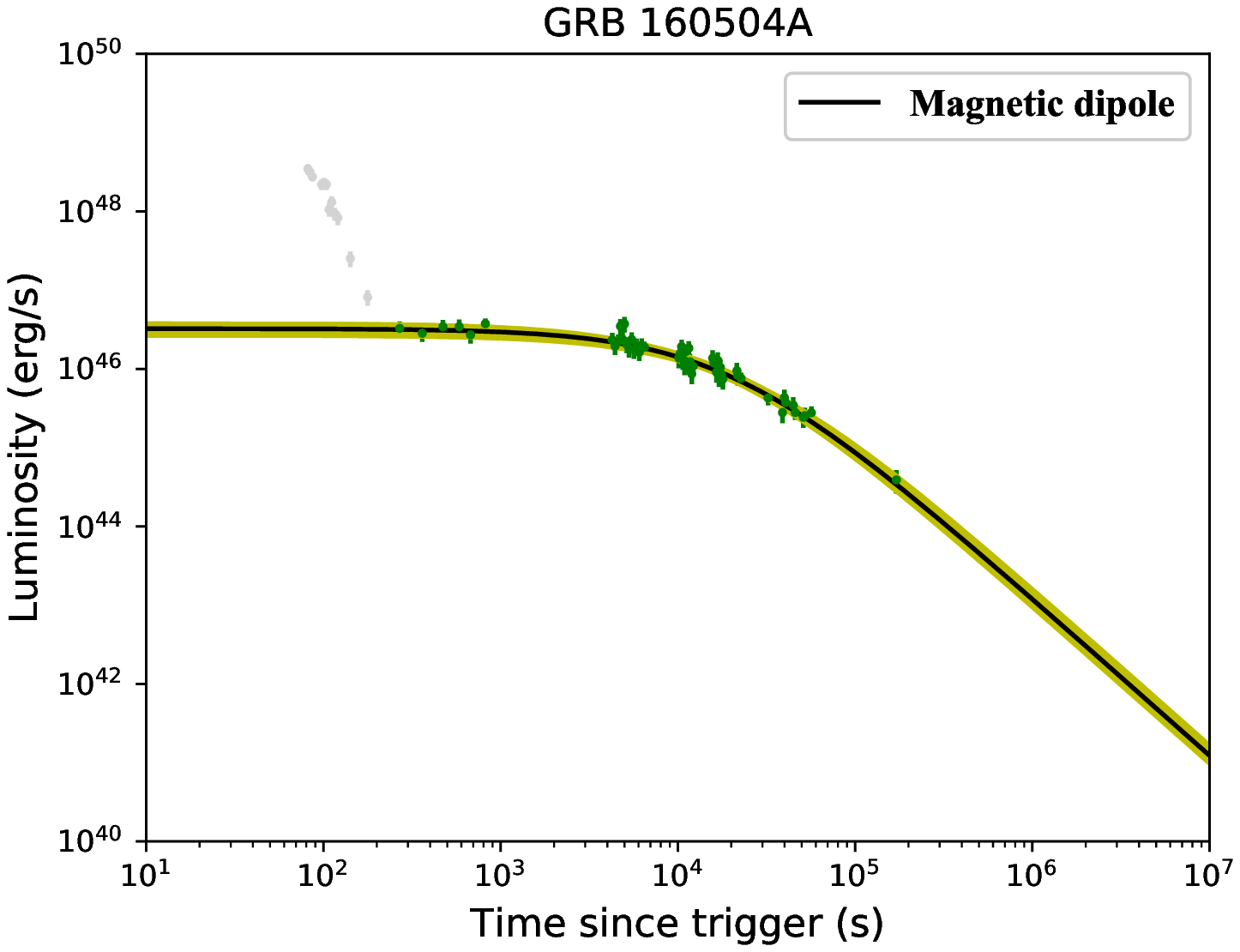}\hspace{-6mm}
\includegraphics[width=0.35\textwidth]{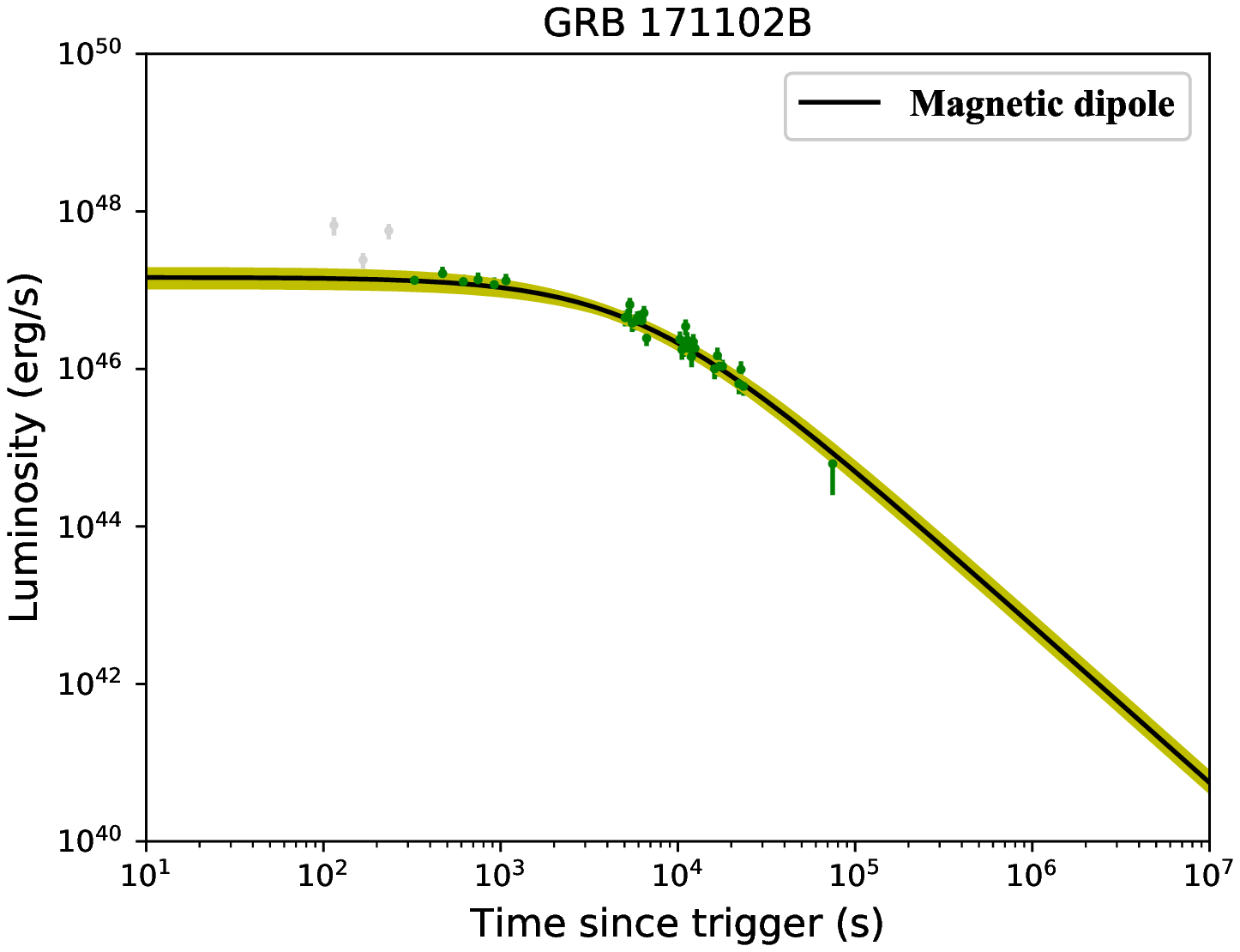}\hspace{-6mm}
\includegraphics[width=0.35\textwidth]{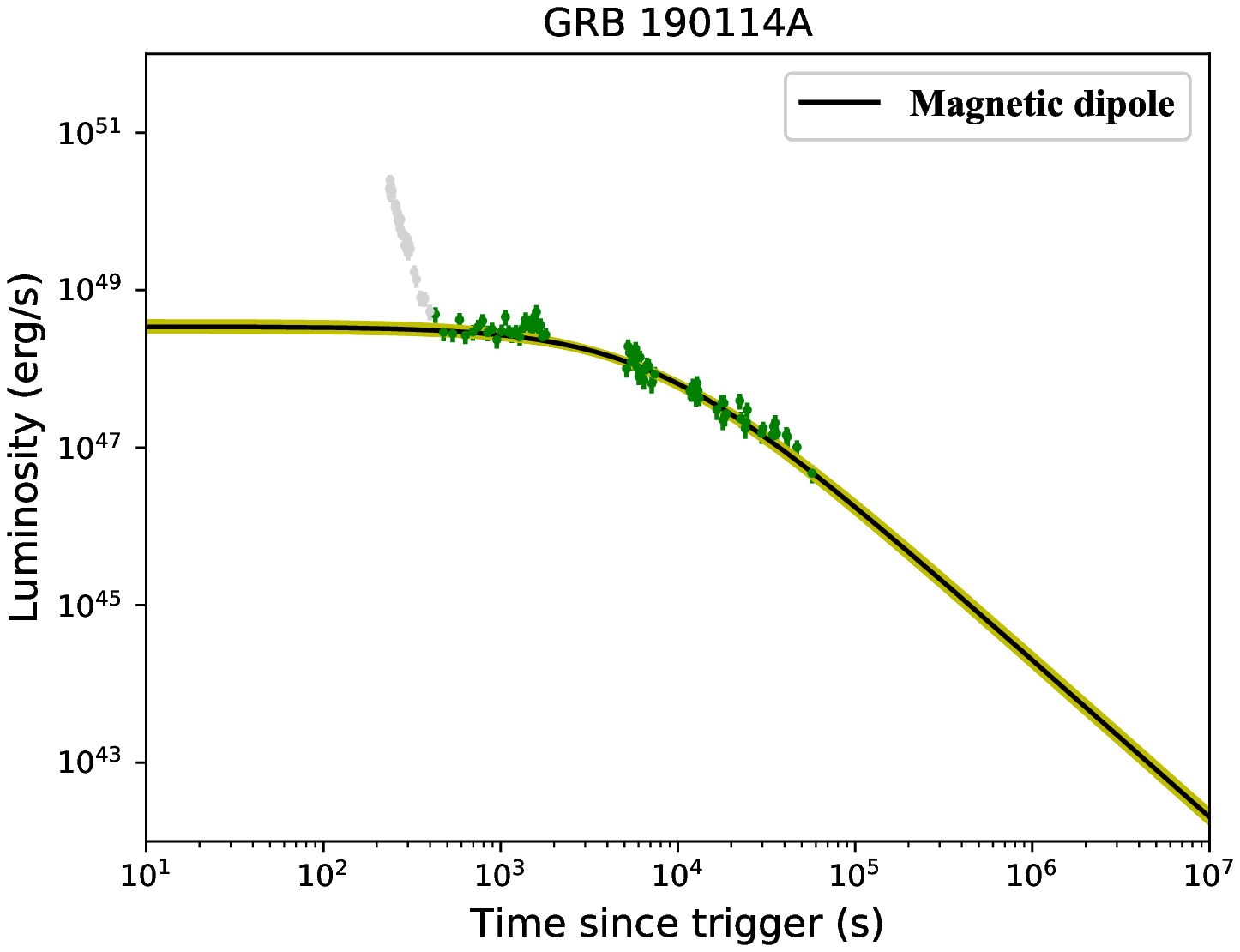}\hspace{-6mm}
\includegraphics[width=0.35\textwidth]{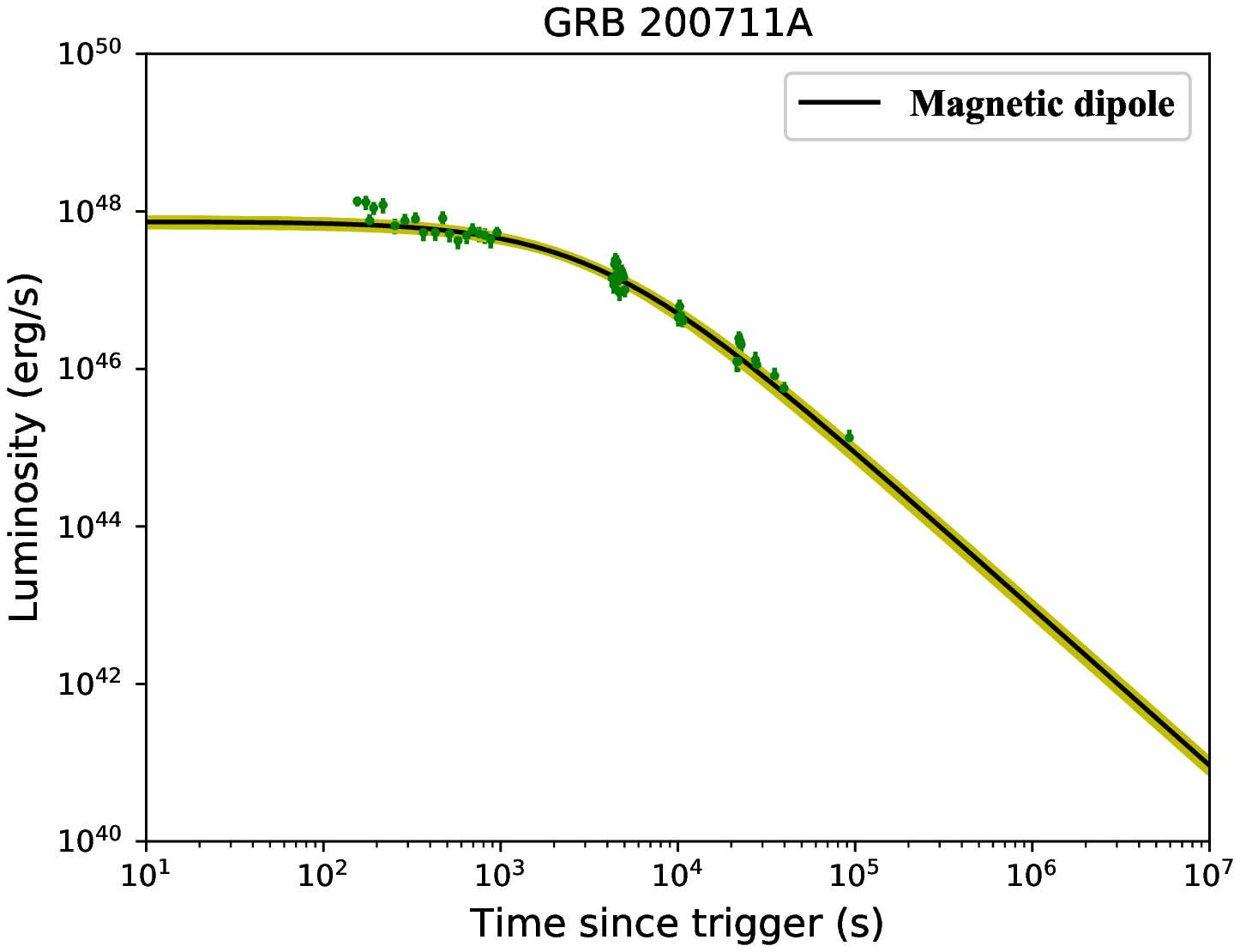}\hspace{-6mm}
\includegraphics[width=0.35\textwidth]{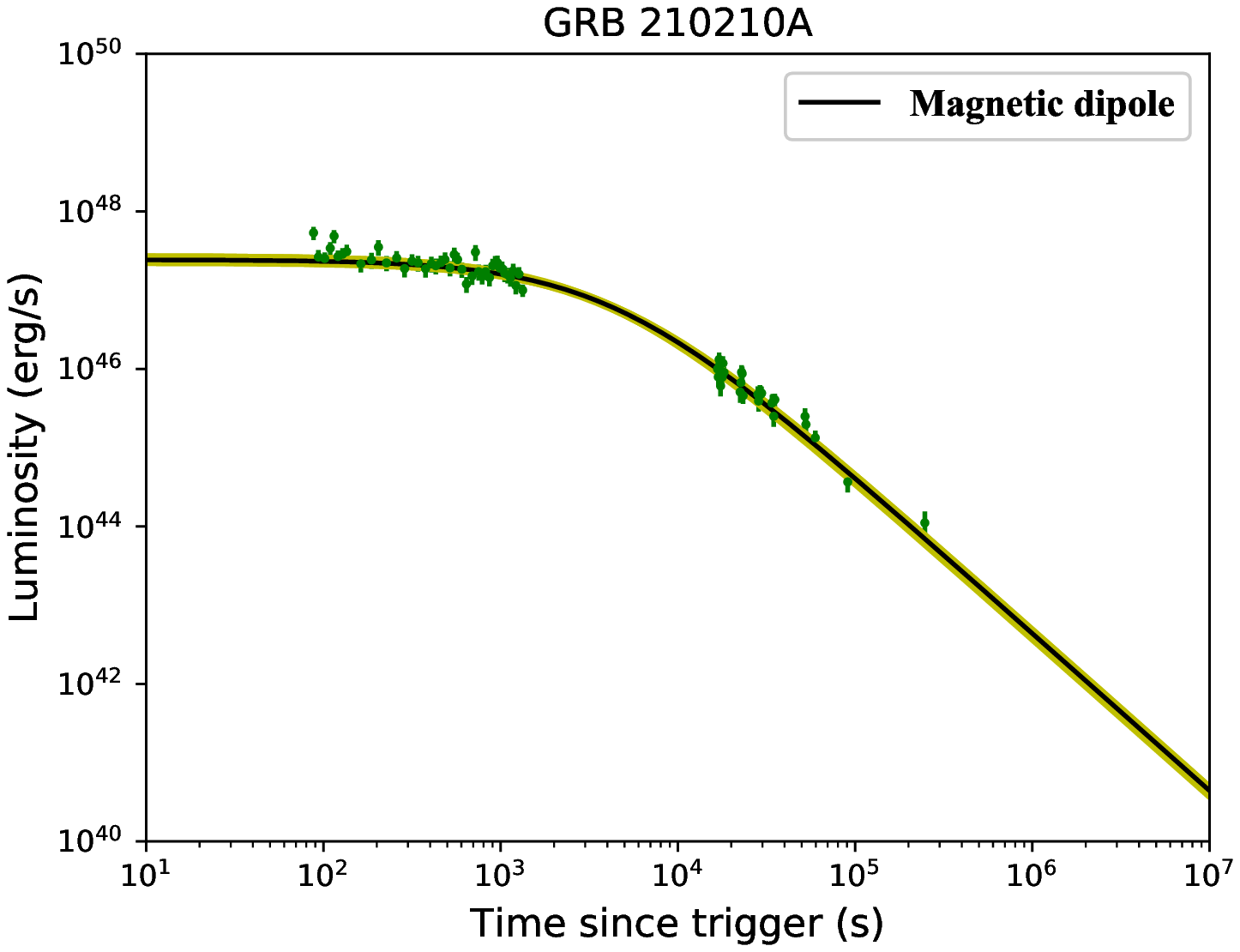}\hspace{-6mm}

\caption {15 LGRB XRT light curves which are well fitted with the MD model. The green data points are the XRT light curves of LGRBs. The black curves show the best-fitting results for the MD model.
The yellow band is the superposition of 200 predicted curves randomly selected from posterior distribution.}
\label{fig-MDsamplexrt}
\end{figure*}

Fig.\ref{fig-GWsamplexrt} and Fig.\ref{fig-MDsamplexrt} show the X-ray light curves of 30 LGRBs, in which the X-ray plateau emission extending to thousands of seconds and followed by a decay segment are present.
These characteristics of light curves are consistent with the expectation of the magnetar model.
We then fit these X-ray light curves with our magnetar model.
Fig.\ref{fig-GWsamplexrt} shows that 15 LGRBs are better fitted with the hybrid model compared to the MD and GW model.
The best-fitting parameters of magnetars are reported in Table \ref{tab:LGRBexample_table}.
The decay slope after the X-ray plateau is related to the braking index.
We constrain the braking index $n$ of 15 LGRBs by fitting their X-ray light curves with equation $ L(t)\propto(1+\frac{t}{\tau})^{\frac{4}{1-n}}$ \citep{L2019},
which is obtained by integrating the torque equation $\dot{\Omega}=-k\Omega^{n}$ \citep{Lasky2017}.
We derive the values of the braking index and find they are in the range of 3 to 5, as shown in Table \ref{tab:LGRBexample_table}, which strongly indicates that the spin-down luminosity of these 15 LGRBs may include the contribution of GW radiation.
Fig.\ref{fig-MDsamplexrt} shows other 15 LGRBs that are better fitted with the MD model.
It is worth noting that no candidate for GW-dominated case has been found in the light curves of our sample.

\begin{figure}
 \centering
\includegraphics[angle=0,scale=0.6]{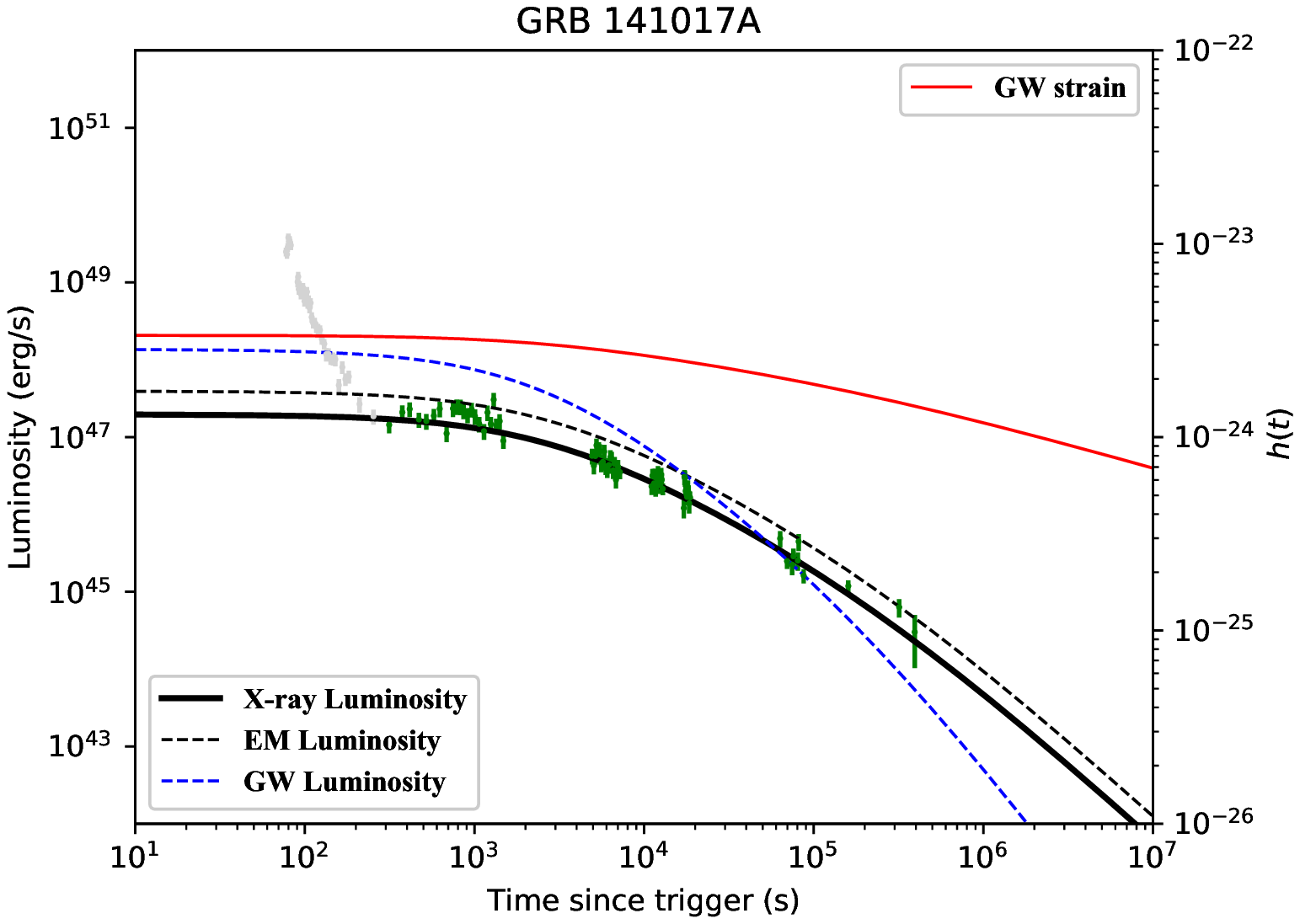}
\includegraphics[angle=0,scale=0.5]{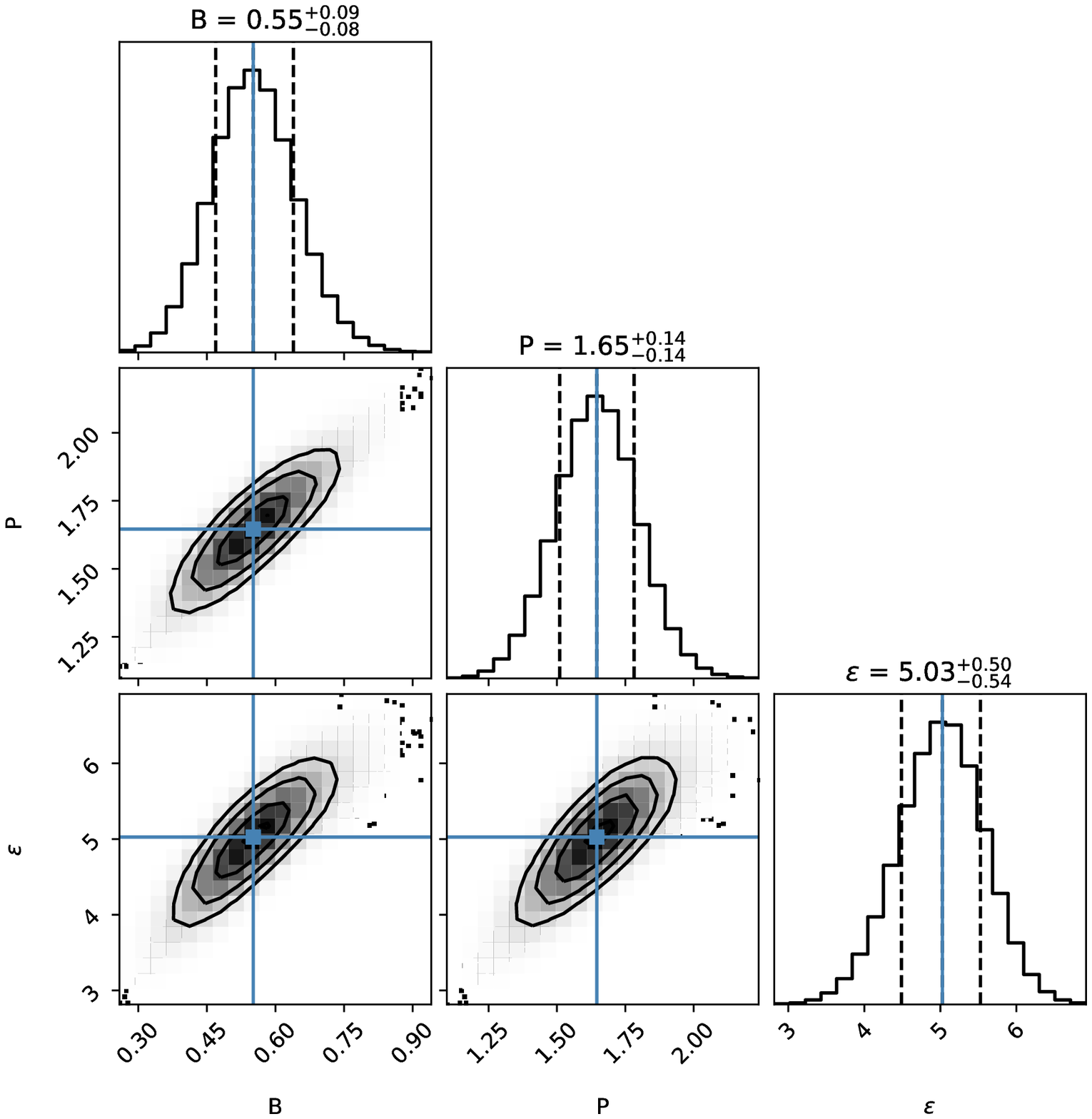}
\caption {Top: The time evolution of X-ray luminosity (black solid line), MD luminosity (black dashed line), GW luminosity (blue dashed line) and GW strain (red solid line) of GRB 141017A. Bottom: The corner plots of the GRB 141017A. The vertical dashed lines represent the $1\sigma$ confidence level of the parameters. The magnetic field $B_p$, period $P_0 $ and ellipticity$\epsilon$ are in units of $10^{15}\mbox{ G}$, $1\mbox{ ms}$ and $10^{-3}$, respectively.}
\label{fig-141017A}
\end{figure}

In the case of GRB~141017A, its spin-down light curve shows a transition from the GW-dominated stage to the MD-dominated stage, which is best fitted by the hybrid model.
As an example, we show the time evolution of X-ray luminosity $L_{\rm X}$, EM luminosity $L_{\rm EM}$ and GW luminosity $L_{\rm GW}$ of GRB~141017A in Fig.\ref{fig-141017A}.
The evolution of GW luminosity can be inferred by combining the equations (\ref{eq:spin}) and (\ref{eq:omegadot}), and this luminosity also exhibits a plateau feature before decay.
One can see that the magnetar spin-down was dominated by GW emission in the early stage and then by MD emission in the late stage.
The corner plots of the GRB~141017A are shown in Fig.\ref{fig-141017A}.

\section{ Constraining the properties of the new-born magnetar  }

Assuming that a rapidly rotating NS (millisecond magnetar) would remain after the explosion of the GRB \citep{Bucciantini2009},
the MD radiation from the magnetar could generate a Poynting-flux-dominated outflow which can be dissipated by shock collision or magnetic reconnection to power the X-ray plateau.
The comparison between the observed spin-down light curves and the model allows us to constrain the initial spin period $P_0$, diploe magnetic field strength $B_p$ and the ellipticity $\epsilon$ of NS.
As shown in Table \ref{tab:LGRBexample_table}, magnetars with $P_0\sim 1\mbox{ ms}$ and $B_p\sim 10^{14}-10^{15}\mbox{ G}$ usually have the ellipticity $\epsilon \sim 10^{-3}$.
Theoretically, the minimum rotation period of the magnetar is $\sim 0.3-0.5 \mbox{ ms}$ \citep{Cook1994, Koranda1997, Haensel1999}.

\begin{figure}
 \centering
\includegraphics[angle=0,scale=0.3]{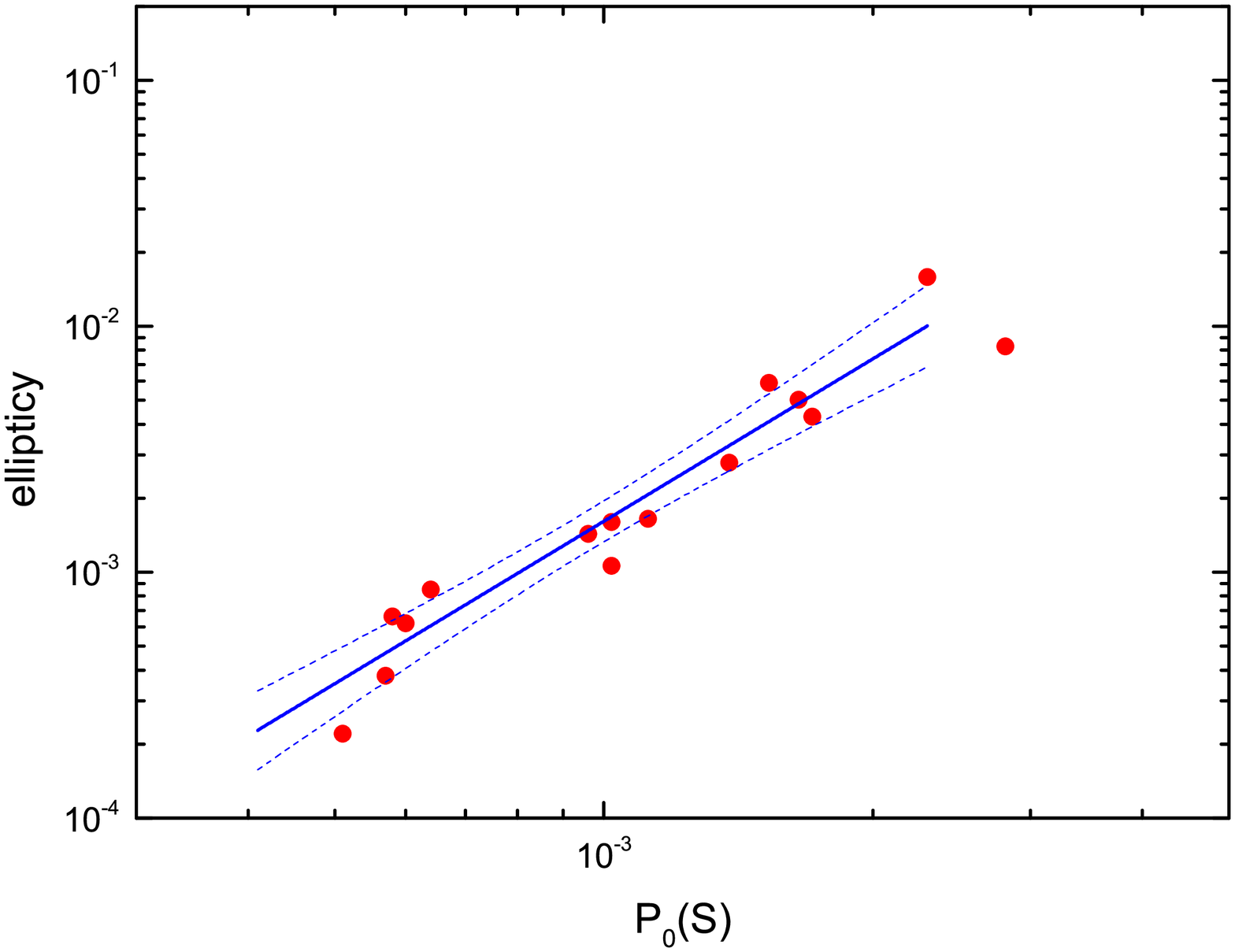}
\includegraphics[angle=0,scale=0.3]{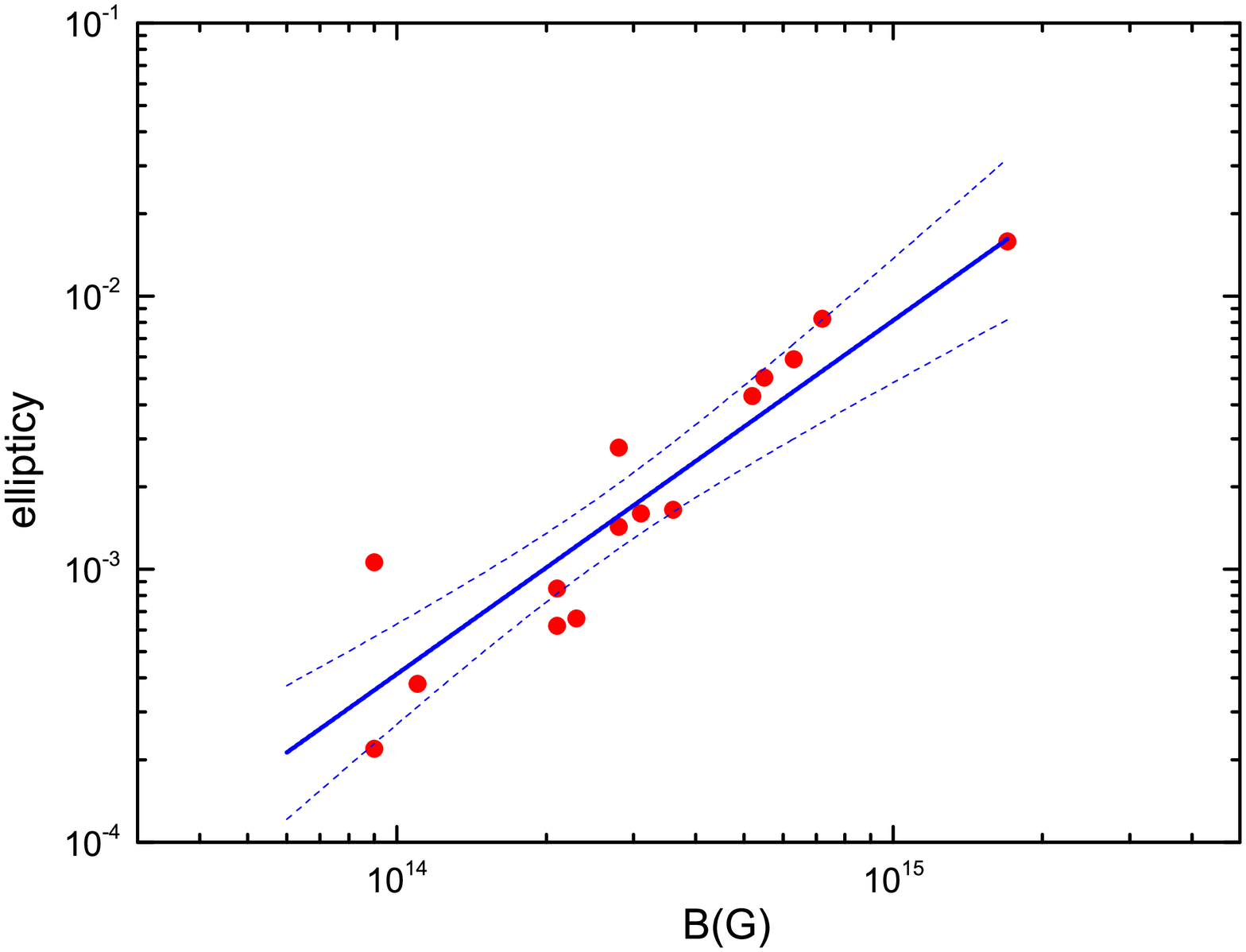}
\includegraphics[angle=0,scale=0.3]{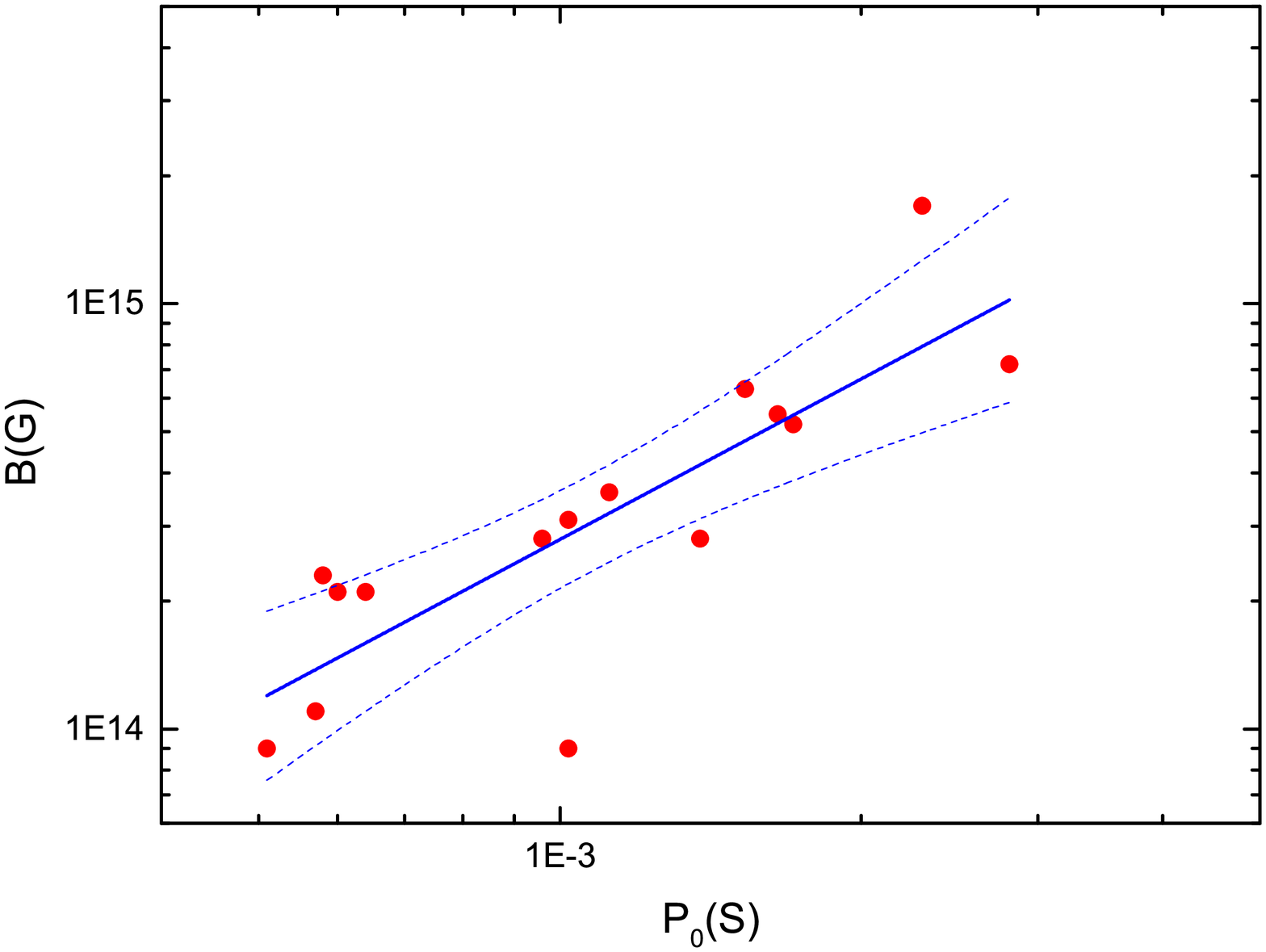}
\caption {Correlation between the $P_0$-$\epsilon$, $ B_p$-$\epsilon$ and $ B_p$-$P_0$ ,respectively. The blue solid lines are the best fitting results, and blue dashed lines are the $95\%$ confidence level.}
\label{fig-B-p-e}
\end{figure}

We perform the ordinary least-squares to estimate the scaling relations of magnetar parameters.
Fig.\ref{fig-B-p-e} shows the distributions of the $\log P_0$-$\log \epsilon$, $\log B_p$-$\log \epsilon$ and $\log B_p$-$\log P_0$, respectively.
The best-fitting relations between the initial spin period $P_0$ and the ellipticity $\epsilon$, diploe magnetic field strength $B_p$ and the ellipticity $\epsilon$,  are
\begin{equation}
 \log\epsilon=3.79_{-0.43}^{+0.52}+ (2.19_{-0.15}^{+0.17})\log P_0,
\end{equation}
\begin{equation}
\log\epsilon= -22.50_{-2.22}^{+2.15} + (1.29_{-0.14}^{+0.15})\log B_p,
\end{equation}
with the Pearson correlation coefficient of $\kappa_1 = 0.84 $ and $\kappa_2 = 0.98 $, and the chance probability $p_1=6.98 \times 10^{-5}$ and $p_2=9.10\times 10^{-11}$.
The correlations of $\epsilon-P_0$ and $\epsilon-B_p$ suggest that the magnetar with a stronger magnetic field and/or a slower spin period corresponds to the larger ellipticity.
The $\epsilon-B_p$ relation can be simply described as $\epsilon \propto B_p^{1.29}$, implying that the NS deformation is related to the dipole magnetic field to some extent.
Recently, some authors argued that the neutron star deformation may be induced by a strong internal magnetic field $B_{\rm int}$ in the stellar core and derived the relation of $\epsilon-B_{\rm int}$ as \citep{Lander2012, Mastrano2012, Lander2014, de2016, de2017,Abbott2020b}
\begin{equation}
	\epsilon\approx 10^{-8}\left(\frac{B_{\rm int}}{10^{12}\,{\rm G}}\right). \label{eq:epsilon}
\end{equation}
According to equation (\ref{eq:epsilon}), in order to achieve $\epsilon \sim 10^{-3}-10^{-4}$, a very strong internal magnetic field ($B_{\rm int}\sim 10^{16}-10^{17}\mbox{ G}$) is need, implying that the strength of the internal field should be $1-2$ orders of magnitude stronger than the external field ($B_{\rm p}\sim 10^{15}\mbox{ G}$).
Similar conclusions were derived from some other studies related to constraining the strength of the internal field for soft gamma-ray repeater (SGR) \citep{Ioka2001, Stella2005,Corsi2011}.
The recurrence rate and energy release of SGR 1900-14 and SGR 1806-20 provide an interior field estimate of $B_{\rm int} \sim 10^{16}\mbox{ G}$ \citep{Stella2005}. 

It is worth noting that the ellipticity values for millisecond magnetars inferred by the GRB data could be different from the conventional pulsars.  For the post-burst magnetar, we have constrained the ellipticity to be $\epsilon \sim 10^{-3}$.  However, most theories and studies of conventional pulsars suggest that the ellipticity of the pulsars are less than  $10^{-6}$ \citep{McDaniel2013}. 
The difference in ellipticity may be attributed to the internal magnetic field strength.
The internal field strength of conventional pulsar may be much weaker than that of the post-burst magnetar, since the magnetar activities (e.g., soft $\gamma$-ray bursts and X-ray emission) are related to the internal magnetic field \citep{Thompson2001,Dall2022}.

Numerical simulations \citep{Stella2005, Dall'Osso2009} suggested that the NS deformation may be caused by magnetic pressure of the internal purely toroidal fields $B_t$, the ellipticity in this scenario should satisfy $\epsilon \propto B_t^{2}$.
The slope of this relationship is slightly steeper than our results, suggesting that the magnetar deformation may be not only caused by the purely toroidal fields.
More discussions about NS deformation will be presented in Section 6.

Fig.\ref{fig-B-p-e} also shows the relation of $\log B_p-\log P_0$. The best-fitting relation gives
\begin{equation}
\log B_p= 18.21_{-0.64}^{+0.69} + (1.29_{-0.21}^{+0.23})\log P_0,
\end{equation}
with $\kappa_3 = 0.76 $ and $p_3=8.68 \times 10^{-4}$ .
This correlation suggests that a longer rotation period corresponds to a stronger magnetic field.

\begin{figure}
 \centering
\includegraphics[angle=0,scale=0.4]{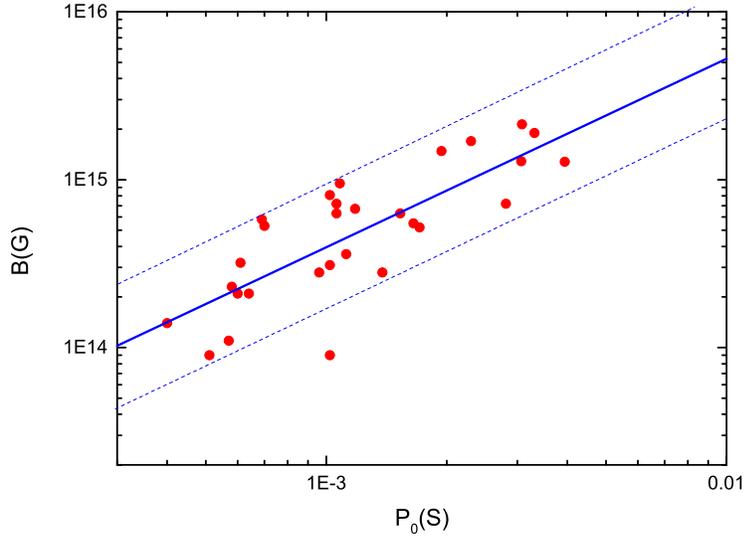}
\caption {$B_p$-$P_0$ distribution for the 30 LGRBs sample. The black solid and dashed lines represent the best fitting result and the $95\%$ confidence level, respectively.}
\label{fig-B-p}
\end{figure}

Adding the other 15 MD-dominated LGRBs to our sample, we derive a more general relation of $\log B_p-\log P_0$ as shown in Fig.\ref{fig-B-p} :
\begin{equation}
\log B_p= 18.01_{-0.46}^{+0.49} + (1.14_{-0.15}^{+0.16})\log P_0,
\end{equation}
with $\kappa_4 = 0.81 $ and $p_4=1.21 \times 10^{-7}$ .
This correlation is highly consistent with the $B\propto P_\mathrm{eq}^{7/6}$ relation for the magnetic propeller model \citep{Piro2011, Gompertz2014}, implying that the MD and GW radiations may occur in the magnetar propeller stage.
By systematically analyzing the GRB X-ray plateaus, \citet{Stratta2018} has deduced similar conclusions.
The interaction between the magnetar and its accretion disk depends on the relative positions of co-rotation radius ($r_\mathrm{c}$), Alfv\'en radius ($r_\mathrm{m}$) and light cylinder radius ($r_\mathrm{L}$).
When the magnetar was in the propeller regime ($r_\mathrm{c}\simeq r_\mathrm{m}$), its spin period would reach an equilibrium state \citep{Piro2011, Lin2020}.
\begin{equation}
P_\mathrm{eq}=2\pi(GM)^{-5/7}R^{18/7}B^{6/7}\dot{M}^{-3/7}.
\label{eq: Peq}
\end{equation}
where $\dot{M}$ is the accretion rate of magnetar.
For the parameters of magnetar inferred from the LGRB data, we can estimate the accretion rate as $\dot{M} \sim 10^{-7}-10^{-2}$ M$_\odot$ s$^{-1}$.

\section{ Detectability of GW signal from magnetar }

The millisecond magnetars formed from the core-collapse of massive stars have been thought to be the potential sources of
continuous GW emission for the Advanced LIGO and Virgo detectors.  Rotating magnetars with asymmetrical deformation would emit observable GWs associated with the GRB X-ray plateau.
Such deformation can be created by the magnetic pressure of the internal magnetic field, or through the
excitation of fluid oscillation.
A triaxial body possessing the mass quadrupole can emit a characteristic gravitational
wave strain \citep{Corsi2009, Howell2011}
\begin{eqnarray}
h(t)=\frac{4G\Omega^2}{c^4d}I\epsilon,
\label{eq:ht}
\end{eqnarray}
where $d$ is the distance to the source.

The optimal matched filter signal-to-noise ratio is defined by \citep{Corsi2009}
\begin{eqnarray}
	\nonumber \rho^2_{max}=\int^{+\infty}_{0}\frac{f^2 h^2(t)(dt/df)}{f S_h(f)} d(\ln f)\\=\int^{+\infty}_{0} \left(\frac{h_c}{h_{rms}}\right)^{2} d(\ln f),
	\label{eq:rho}
\end{eqnarray}
where $h_{rms}=\sqrt{f S_h(f)}$ is the detector noise curve, $S_h(f)$ is the power spectral density of the detector noise, $h_c=f h(t)\sqrt{dt/df}$ is the characteristic amplitude of GW signal.
For magnetars with significant GW emission in spin-down process, $h_c$ can be written as \citep{Corsi2009, Howell2011, Fan2013}
\begin{equation}
h_c= \frac{1}{d}\sqrt{\frac{5GIf}{2c^3}},
\label{eq:hc}
\end{equation}
where $f=\Omega/\pi$ is the GW frequency.

It can be inferred from equations (\ref{eq:ht}) and (\ref{eq:hc}) that GW strain depends on the evolution of stellar angular frequency and distance to the source. The evolution of angular frequency can be inferred by equation (\ref{eq:omegadot}), therefore, one can derive the GW strain by using the observed spin-down light curves. Fig.\ref{fig-141017A} shows the GW strain of GRB 141017A, which exhibits a plateau segment then followed by slowly decay..

\begin{figure}
\centering
 \includegraphics[width=0.6\textwidth]{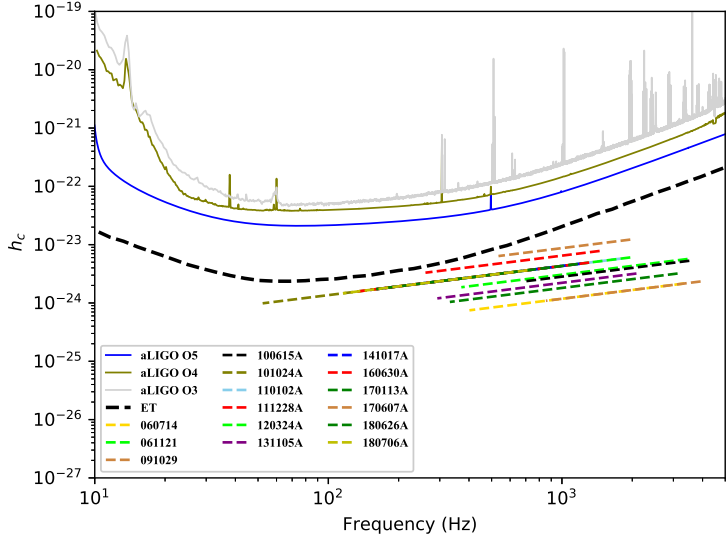}
 \caption {GW amplitude evolution for the 15 LGRBs with GW radiation contribution. The dashed black curves represent the projected sensitivity for ET. The light grey curves, olive curves and blue curves are the O3, O4 and O5 design sensitivity for aLIGO, respectively. }
 \label{fig-hc}
\end{figure}

We calculated  the GW amplitude of 15 LGRBs with significant GW radiation as well as the projected sensitivity for aLIGO and ET detectors in Fig.\ref{fig-hc}\citep{Abbott2018}.
Comparing with the sensitivities for aLIGO and ET detectors, one can find that GW signals from these LGRBs may not reach the sensitivity threshold of the aLIGO and ET detectors.
From equation (\ref{eq:hc}), we can estimate the detection horizon of the magnetar for the aLIGO.
For a rapidly spinning magnetar ($P\sim 1\mbox{ ms}$, $ B \sim10^{15}\mbox{ G}$ ), the detection horizons for aLIGO O5 is $\sim 180\mbox{ Mpc}$.
If the GW signal associated with the X-ray plateau of magnetar can be detected in the future, it would directly prove that magnetar can act as the central engine of GRBs.

\section{Conclusion and Discussion}

A rapidly rotating magnetar may survive after a GRB explosion.
Under the assumption of the magnetar model, we analyzed the X-ray light curves of 30 LGRB which show an X-ray plateau emission and followed by a decay segment.
New-born magnetar may undergo non-axisymmetric deformation or various stellar oscillations, which could emit the continuous GW emission associated with the X-ray plateau.
We try to search for GW emission signature from a sample of LGRBs by measuring the plateau and decay index.
We utilize the X-ray observations of the LGRBs to constrain the properties of the new-born magnetar, including the initial spin period, dipole magnetic field strength and ellipticity.
Moreover, we derive the GW strain for magnetar sample via their spin-down processes, suggesting that GWs from these objects may not be detectable by the aLIGO and ET detectors.
For a rapidly spinning magnetar ($P\sim 1\mbox{ ms}$, $ B \sim10^{15}\mbox{ G}$ ), the detection horizons for aLIGO O5 is $\sim 180\mbox{ Mpc}$.

Deriving the parameters of magnetar by modeling the LGRB X-ray plateau light curves, we find some tight relations between magnetar parameters, e.g.,
$\epsilon \propto B_p^{1.29}$ and $B_p\propto P_0^{1.14}$.
The correlations of $\epsilon-P_0$ and $\epsilon-B_p$ suggest that the magnetar with a stronger magnetic field and/or a slower spin period corresponds to the larger ellipticity. The relation of $\log B_p-\log P_0$ indicates the MD and GW radiation of magnetar may occur in the magnetic propeller phase as it spins down.
We use the magnetar model to measure the ellipticity of NS. The ellipticity for most LGRBs is constrained to be about $10^{-3}$, implying that magnetar would lose significant rotation energy via GW emission if the ellipticity of magnetar is larger than $10^{-3}$.
By using the statistical properties of SGRBs, \citet{Gao2016} suggested that only the ellipticity and dipole field strength of magnetars are around $\epsilon \sim 5\times10^{-3}$ and $ B \sim10^{15}\mbox{ G}$ can reproduce the distributions of GRB X-ray plateau and duration.
\citet{Lasky2016} presented the different ways of inducing NS deformation and constrained the corresponding ellipticity.
The values of ellipticity inferred by our LGRB sample are consistent with the results suggested by \citet{Gao2016} and \citet{Lasky2016}.

The ellipticity of magnetar depends on the deformation mechanisms.
If the magnetic field induces NS deformation, several scenarios can be
described as follows: a newly born magnetar formed in the collapse of a massive star is differentially rotating. The internal field could be amplified to result in a strong toroidal magnetic field due to differential rotation and the magnetic rotation instability. The purely toroidal magnetic field may induce NS deformation to produce the large ellipticity, e.g., $\epsilon\approx 0.016 \left({B_{\rm t}}/{10^{17}\,{\rm G}}\right)^2$ \citep{Stella2005, Dall'Osso2009}.
Alternatively, if the poloidal field dominates and induces the NS deformation, the ellipticity is
$\epsilon\approx B_\mathrm{pole}^{2}/\left({\pi{\mu_0}G{\rho^2} {R^2}}\right)$ \citep{Bonazzola1996, Konno2000}.
The relation of $\epsilon \propto B_p^{1.29}$ inferred by our sample is not consistent with the theoretical relations predicted by the purely poloidal magnetic field and the purely toroidal magnetic field.  Our results suggest that magnetar deformation may be induced by a disordered magnetic field composed of a strong mixed toroidal-poloidal field \citep{Thompson2002}.

The relation of $\log B_p-\log P_0$ is consistent with the $B\propto P_\mathrm{eq}^{7/6}$ relation for magnetic propeller model.
In the magnetic propeller phase, magnetar may reach the equilibrium spin period due to material fall-back accretion \citep{Lin2020}.
The matter at the edge of the accreting disk would flow into the poles of the magnetar along with the magnetic field lines and then would form two accreting columns (known as ``mountain'' ). In this scenario, magnetar has a rapid time-varying quadrupole moment, allowing it to become a source of GW emission.
The ellipticity of this source depends on the mass of accretion \citep{Haskell2015, Zhong2019, Sur2021}.
Therefore, a ``mountain'' from accreting magnetar may be  another potential way for the magnetar to generate GW radiation.

\section*{ Acknowledgements }
We gratefully thank the anonymous referee for helpful comments to improve this paper. This work made use of data supplied by the UK Swift Science Data Centre at the University of Leicester. This work was supported by NSFC (No. 12073080, 11933010, 11921003) and by the Chinese Academy
of Sciences via the Key Research Program of Frontier Sciences (No. QYZDJ-SSW-SYS024).

\bibliography{bibtex}
\clearpage

\begin{deluxetable}{ccccccccccccccccc}
\tablewidth{0pt} \tabletypesize{\small} \tablecaption{  Constraints on properties of the new-born magnetar in our LGRB sample }
\tablehead{ \colhead{GRB}& \colhead{Redshift}& \colhead{$L_{p}\tablenotemark{a}$ ($\rm erg~s^{-1}$)}& \colhead{$B_{p}$ ($\times 10^{15}$ G)}&
\colhead{$P_{0}$ (ms) }& \colhead{$\epsilon$ }& \colhead{$n\tablenotemark{b}$ }&  \colhead{$\chi^{2}$/dof}}\startdata
     060714	    &	2.71 &$\sim1.64e48$&	 $0.21^{+0.04}_{-0.03}$&    $0.60^{+0.05}_{-0.05}$ 	&	$0.62e-3$	&$3.86^{+0.12}_{-0.13}$	    &$80/46$	\\
     061121	    &	1.31 &$\sim2.25e48$ &    $0.23^{+0.01}_{-0.01}$&	$0.58^{+0.01}_{-0.01}$	&	$0.66e-3$	&$3.67^{+0.03}_{-0.03}$     &$514/279$	\\
     091029	    &	2.75 &$\sim5.77e47$ &    $0.09^{+0.05}_{-0.04}$&	$0.51^{+0.01}_{-0.01}$ 	&	$0.22e-3$	&$4.11^{+0.10}_{-0.10}$	    &$252/119$	\\
     100615A	&	1.39 &$\sim5.53e47$ &    $0.11^{+0.02}_{-0.02}$ &	$0.57^{+0.06}_{-0.05}$  &   $0.38e-3$	&$4.44^{+0.21}_{-0.22}$	    &$150/81$	\\
     101024A	&	     &$\sim4.97e47$ &    $1.70^{+0.24}_{-0.23}$ &	$2.30^{+0.18}_{-0.18}$ 	&	$1.59e-2$	&$3.67^{+0.03}_{-0.03}$	    &$82/56$	\\
     110102A	&	     &$\sim3.97e47$ &    $0.36^{+0.02}_{-0.02}$ &	$1.12^{+0.04}_{-0.04}$ &	$1.65e-3$	&$3.65^{+0.06}_{-0.06}$	    &$412/269$	\\
     111228A	&	0.72 &$\sim1.04e47$ &    $0.28^{+0.03}_{-0.03}$ &	$1.38^{+0.07}_{-0.07}$ &	$2.79e-3$	&$4.09^{+0.07}_{-0.07}$	    &$286/143$	\\
     120324A	&	     &$\sim4.28e47$ &    $0.31^{+0.04}_{-0.04}$ &	$1.02^{+0.04}_{-0.04}$ &	$1.60e-3$	&$4.02^{+0.13}_{-0.13}$     &$254/145$	\\
     131105A   &	1.69 &$\sim4.45e47$ &    $0.28^{+0.09}_{-0.08}$ &   $0.96^{+0.15}_{-0.15}$  &	$1.43e-3$	&$4.05^{+0.21}_{-0.22}$	    &$81/51$	\\
     141017A	&	     &$\sim1.97e47$ &    $0.55^{+0.09}_{-0.08}$ &	$1.65^{+0.23}_{-0.23}$ &	$5.03e-3$	&$4.07^{+0.13}_{-0.13}$	    &$166/77$	\\
     160630A	&	     &$\sim3.49e47$ &    $0.63^{+0.19}_{-0.17}$ &	$1.53^{+0.05}_{-0.05}$ &	$5.88e-3$	&$4.18^{+0.19}_{-0.19}$	    &$153/66$	\\
     170113A	&	1.97 &$\sim2.53e48$ &    $0.21^{+0.02}_{-0.02}$ &	$0.64^{+0.04}_{-0.04}$ &	$0.85e-3$	&$4.01^{+0.08}_{-0.08}$     &$272/127$	\\
     170607A	&	0.56 &$\sim3.61e46$ &    $0.09^{+0.03}_{-0.03}$ &	$1.02^{+0.15}_{-0.18}$ &	$1.06e-3$	&$4.76^{+0.11}_{-0.11}$     &$389/171$	\\
     180626A	&	     &$\sim1.52e47$ &    $0.52^{+0.08}_{-0.07}$ &	$1.71^{+0.15}_{-0.14}$ &	$4.30e-3$	&$3.95^{+0.14}_{-0.14}$	    &$141/80$	\\
     180706A	&	0.30 &$\sim4.01e46$ &    $0.72^{+0.32}_{-0.27}$ &	$2.81^{+0.61}_{-0.60}$ &	$8.28e-3$	&$4.71^{+0.56}_{-0.55}$     &$33/38$
\enddata
\tablenotetext{a}{The X-ray plateau luminosity in our sample.}
\tablenotetext{b}{The braking index of magnetar.}
\label{tab:LGRBexample_table}
\end{deluxetable}

%% This command is needed to show the entire author+affiliation list when
%% the collaboration and author truncation commands are used.  It has to
%% go at the end of the manuscript.
%\allauthors

%% Include this line if you are using the \added, \replaced, \deleted
%% commands to see a summary list of all changes at the end of the article.
%\listofchanges

\end{document}